\documentclass[superscriptaddress,twocolumn]{revtex4}
\usepackage{amsmath}
\usepackage{amssymb}
\usepackage{graphicx}
\usepackage{hyperref}
\usepackage{amsfonts}

\usepackage{doi}
\usepackage{xcolor}
\begin{document}
\title{Active elastocapillarity in soft solids with negative surface tension}
\author{Jack Binysh}
\affiliation{Department of Physics, University of Bath, Claverton Down, Bath, BA2 7AY, UK}
\author{Thomas R.~Wilks}
\affiliation{School of Chemistry, University of Birmingham, Edgbaston, Birmingham B15 2TT, UK}
\author{Anton Souslov}
\email{a.souslov@bath.ac.uk}
\affiliation{Department of Physics, University of Bath, Claverton Down, Bath, BA2 7AY, UK}
\begin{abstract}
Active solids consume energy to allow for actuation, shape change, and wave propagation not possible in equilibrium. Whereas active interfaces have been realized across many experimental systems, control of three-dimensional (3D) bulk materials remains a challenge. Here, we develop continuum theory and microscopic simulations that describe a 3D soft solid whose boundary experiences active surface stresses. The competition between active boundary and elastic bulk yields a broad range of previously unexplored phenomena, which are demonstrations of so-called active elastocapillarity. In contrast to thin shells and vesicles, we discover that bulk 3D elasticity controls snap-through transitions between different anisotropic shapes. These transitions meet at a critical point, allowing a universal classification via Landau theory. The active surface modifies elastic wave propagation to allow zero, or even negative, group velocities. These phenomena offer robust principles for programming shape change and functionality into active solids, from robotic metamaterials down to shape-shifting nanoparticles.
\end{abstract}

\maketitle
\section{Introduction}
Embedding stress-generating active elements into a passive solid powers functionality inaccessible in thermal equilibrium. These active metamaterials occupy the space between materials and machines, giving rise to exotic phenomena from actuation and shape change~\cite{Banerjee_2011, maitra_oriented_2019, prost_active_2015, mietke_minimal_2019,  liu_viscoelastic_2021, marchetti_hydrodynamics_2013,ronceray_stress-dependent_2019, mizuno_nonequilibrium_2007, woodhouse_autonomous_2018, miller_geometry_2018, hawkes_programmable_2010, santangelo_extreme_2017, hua_anisotropic_2019} to overdamped wave propagation~\cite{scheibner_odd_2020, banerjee_active_2020,raney_stable_2016,nadkarni_unidirectional_2016} and nonreciprocal interactions~\cite{brandenbourger_non-reciprocal_2019, braverman_topological_2020,gupta_active_2020}. For two-dimensional~(2D) active interfaces, powerful design principles exist for the distribution and control of stress-generating elements in order to achieve a target behavior~\cite{hawkes_programmable_2010, santangelo_extreme_2017, keber_topology_2014, salbreux_mechanics_2017,pearce_programming_2020,mostajeran_frame_2017}. A key challenge is to develop such principles for the control of bulk three-dimensional~(3D) solids. Realizations of far-from-equilibrium solids range from macroscopic mechanical materials~\cite{woodhouse_autonomous_2018,brandenbourger_non-reciprocal_2019} and hydrogels~\cite{sato_matsuo_kinetics_1988,chang_extreme_2018,zhang_non-uniform_2019} down to the microscale~\cite{prost_active_2015,liu_viscoelastic_2021,pearce_programming_2020, hua_anisotropic_2019}. The challenge of spanning these systems and length scales requires principles based on a continuum approach.

In thermal equilibrium, the shape and structure of a soft solid is determined not only by 3D bulk elasticity, but also by surface stresses on its 2D boundary. This competition, termed (passive) elastocapillarity~\cite{style_elastocapillarity_2017, bico_elastocapillarity_2018}, has been used to stiffen composites~\cite{style_surface_2015}, self-assemble micro-objects \cite{py_capillary_2007}, and drive the coiling of nanoparticle helices~\cite{pham_highly_2013}. These phenomena all originate in the minimization of surface area due to the isotropic surface stress tensor $\Upsilon^p_{ij}=\gamma_p \delta_{ij}$, where $\delta_{ij}$ is the Kronecker delta. Because passive elastocapillary solids are in equilibrium, their surface tension $\gamma_\mathrm{p}$ is constrained to be positive.

In this work, we show that elastocapillarity offers a distinct set of design principles when considering 3D soft solids driven out of equilibrium. The interplay between bulk and surface stresses translates to this far-from-equilibrium context, but now each stress tensor may acquire additional, active terms~\cite{salbreux_mechanics_2017,scheibner_odd_2020}. We refer to the resulting phenomenology as active elastocapillarity. This scenario stands in contrast to the buckling of elastic shells 
~\cite{landau_theory_1986,quilliet_anisotropic_2008,hannezo_theory_2014, mietke_minimal_2019,kusters_actin_2019}, or morphing of thin programmable sheets~\cite{mostajeran_frame_2017,van_rees_growth_2017,griniasty_curved_2019, pearce_programming_2020}. Rather than a thin layer, we consider a fully 3D solid,~Fig.~\ref{fig:examples}{\bf a}, which is sufficiently soft for active surface stresses to strain the entire bulk. We focus on a minimal change to the surface stress tensor due to activity. This is the addition of an isotropic dilational stress $\Upsilon^a_{ij}\equiv\gamma_a \delta_{ij}$, where $\gamma_a <0$. Sufficiently strong dilations will overpower passive tension $\gamma_p$, giving an overall surface tension $\gamma$ which is effectively negative, $\gamma\equiv\gamma_\mathrm{a} + \gamma_\mathrm{p} < 0$~\cite{patashinski_unstable_2012,kusters_actin_2019,turlier_unveiling_2019}. 

Intuitively, positive surface tension rounds any 3D object into a sphere. For negative surface tension, is there a unique favoured shape? Developing a continuum theory of active elastocapillarity, here we show that final shape is not determined by $|\gamma|$ alone. Instead, it can be selected by varying either surface stresses or bulk elastic moduli. As active driving increases, we find that distinct geometries discontinuously snap between one another, as realized in a simple particle-based numerical model. As well as shape, active surface stresses also control dynamic phenomena. For example, we find that negative surface tension softens Rayleigh wave propagation, leading to zero (or even negative) group velocity. Taken together, our results form a toolkit for programming the functionality of 3D active solids. 

\subsection{Experimental Realizations}

\begin{figure}[t!]
\centering
\includegraphics[width=.9\linewidth]{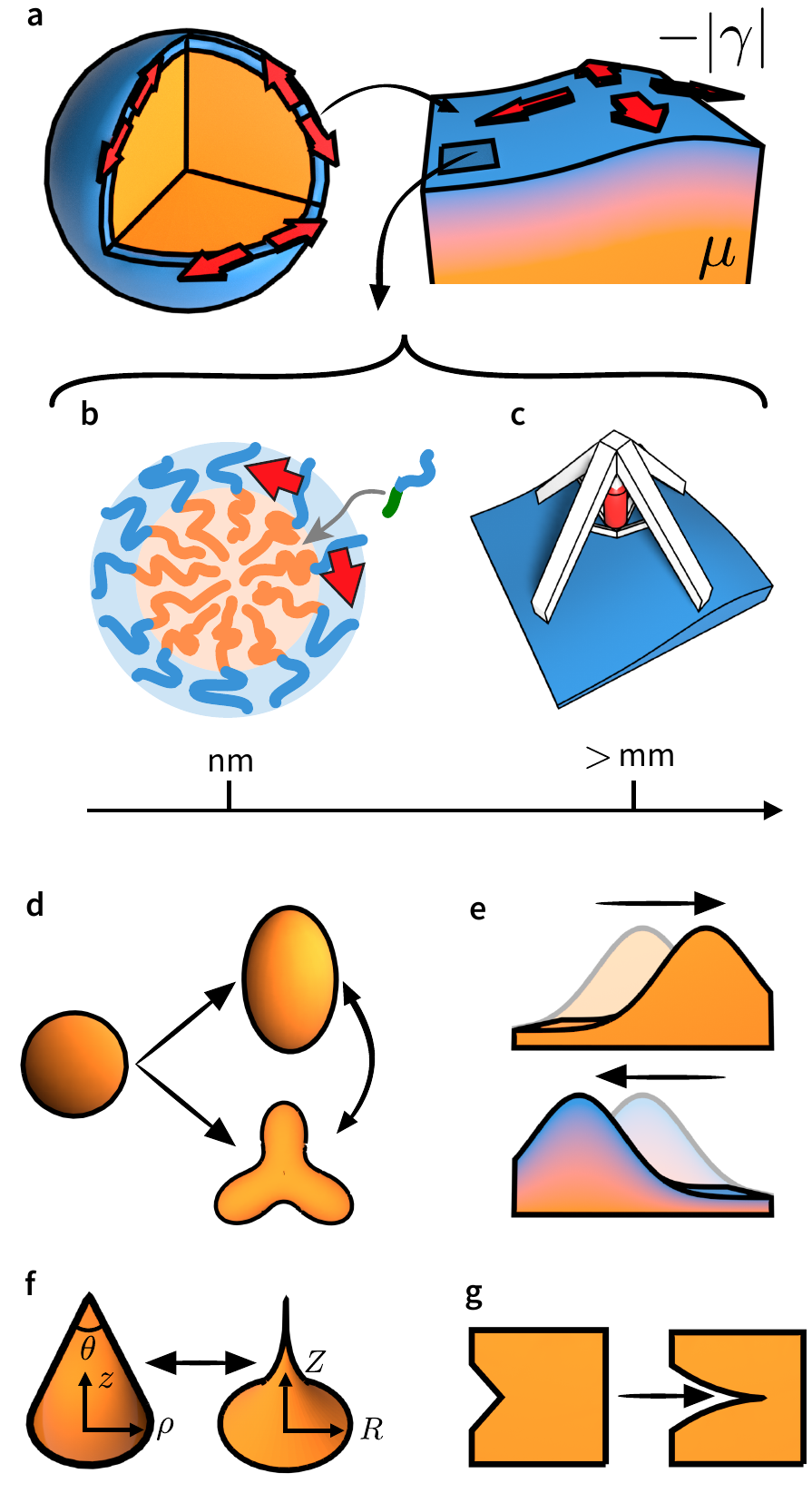}
    \caption{{\bf Active elastocapillary phenomena in soft solids with negative surface tension}. {\bf a}. At the boundary of a bulk 3D soft solid, an active layer (blue) exerts dilational surface stresses (red arrows), which compete against bulk elasticity (orange). Left panel shows a cutaway view of the entire solid, right panel shows a zoom-in near the boundary.~{\bf b}--{\bf c}. Mechanisms leading to active surface stresses, from the nanoscale to the macroscale: {\bf b}.~Insertion of complementary polymers into a nanoparticle surface causes steric crowding~\cite{hua_anisotropic_2019}. {\bf c}.~An active metamaterial in which mechanical actuators are embedded in the surface of an elastic medium~\cite{hawkes_programmable_2010,brandenbourger_non-reciprocal_2019,santangelo_extreme_2017}. {\bf d}--{\bf g}.~Continuum phenomenology: {\bf d}.~Switching between spheres and complex shapes of higher surface area. {\bf e}.~Negative group velocity of elastocapillary Rayleigh waves. {\bf f}.~Sharpening of corners and edges, with a cone of tip angle $\theta$ deforming into a cusp of the form $|Z(R)| \sim \theta^{-1} R^{1/4}$. {\bf g}.~Promotion of crack propagation.
    }
\label{fig:examples}
\end{figure}

In Figs.~\ref{fig:examples}{\bf b--c}, we give examples of experimental mechanisms for how active surface stresses can be designed. At the microscale, Fig.~\ref{fig:examples}{\bf b} shows a nanoparticle made of a diblock copolymer with a hydrophilic head (blue) and a hydrophobic tail (orange). The long tails form a melt, leading to bulk elasticity inside the nanoparticle~\cite{hua_anisotropic_2019}. Insertion of a complementary polymer (blue head, green tail) into the nanoparticle surface drives area growth. As an example, Ref.~\cite{hua_anisotropic_2019} achieves this insertion using binding between paired DNA nucleobases (orange to green). To accomodate this growth, the nanoparticles deform away from their equilibrium spherical shape.

On the macroscale, a simpler realization is to embed dilational motors into the surface of a solid rubber ball (or another object), as shown in Fig.~\ref{fig:examples}{\bf c}. Alternatively, macroscale surface dilation can be designed using self-folding origami sheets or other deployable structures that spontaneously grow their effective surface area as they unfold~\cite{hawkes_programmable_2010,santangelo_extreme_2017}. The incompressibility of the underlying rubber ball prohibits isotropic expansion at fixed spherical shape. Instead the ball shears, with the coupling between rubber shear modulus and dilational surface stresses in the active sheet determining the final shape of the composite object.

We focus on physical realizations using active elements such as independent motors. Like all examples of active matter, these motors locally consume energy to exert mechanical stresses. When these decentralized stresses couple together, shape change can become an emergent phenomenon via spontaneous symmetry breaking. An alternative mechanism for generating negative surface tension commonly used to make wrinkling patterns~\cite{sato_matsuo_kinetics_1988,li_surface_2011, fogle_shape_2013,tallinen_mechanics_2015,budday_size_2015,chang_extreme_2018,kusters_actin_2019} is to globally prestress a planar sheet, which is then glued onto a flat substrate~\cite{chen_herringbone_2004}. By contrast, active elements can lead to a negative surface tension even when embedded into the surface of an arbitrarily shaped object. Such dilational elements are readily found across many lengthscales~\cite{santangelo_extreme_2017,hua_anisotropic_2019}. Here we take a continuum approach which is independent of the microscopic origin of negative surface tension or system scale. For example, in Fig.~\ref{fig:examples}{\bf b}, control over the continuum-level surface tension $\gamma$ can be implemented by varying the concentration of inserted polymer. In Fig.~\ref{fig:examples}{\bf c}, such control can be achieved by varying the forces exerted by the macroscopic motors.
\section{Results}
Figures~\ref{fig:examples}{\bf d--g} illustrate a selection of the active elastocapillary phenomena that arise due to spontaneous growth of surface area.
Below we focus on exact solutions for deformations of a sphere (Fig.~\ref{fig:examples}{\bf d}) as well as surface waves and instabilities (Fig.~\ref{fig:examples}{\bf e}). First, we give an example that typifies the phenomenology of active elastocapillarity: the sharpening of a cone to a cusp (Fig.~\ref{fig:examples}{\bf f}).

Passive elastocapillarity smooths out a vertical cone of tip angle $\theta$~\cite{jagota_surface-tension-induced_2012,liu_energy_2014,mora_softening_2015,mora_solid_2013}. By contrast, active elastocapillarity will sharpen this feature to a power-law cusp. The solid's boundary is then described by the curve $|Z(R)| \sim \theta^{-1} R^{1/4}$, with the deformed conical height $Z$ converging sharply to the origin as a function of its deformed radius $R$. To derive this power law, we consider the elongation of an incompressible elastic cylinder, of elastic modulus $\mu$ and undeformed radius~$\rho$, under the action of negative surface tension on its curved boundary. The stretch factor $\lambda>1$ is found by balancing the elastic deformation energy $\mu \rho^2 \lambda^2$ against the surface energy $|\gamma| \rho \sqrt{\lambda}$ to give $\lambda \sim (|\gamma|/ \mu\rho)^{2/3}$. The deformed height is then $Z=\lambda z$, with a corresponding radial contraction of $R =\rho/\sqrt{\lambda}$. We then take the cone to be a stack of infinitesimally thin cylinders of progressively decreasing radius $\rho\sim\theta z$. Each cylinder experiences a $z$-dependent elongation $\lambda(z) \sim (\theta z)^{-2/3}$. Integrating these extensions, as described in detail in the Methods, and expressing the result in terms of the deformed radius, $R \sim \theta z{\lambda(z)}^{-1/2}\sim \left(\theta z\right)^{4/3}$, yields the $R^{1/4}$ power law. Sharpening of corners corresponds to the case of a small conical angle $\theta$. Taking a large $\theta$ instead corresponds to crack proliferation through stress concentration, shown in  Fig.~\ref{fig:examples}{\bf g}. 

\subsection{Continuum Theory}
We now proceed to the general framework of our continuum description. Active elastocapillarity is defined by two intrinsic length scales. The first, so-called \emph{elastocapillary length} $l_\gamma \equiv |\gamma|/\mu$, measures the ratio of effective surface tension $|\gamma|$ to shear elastic modulus $\mu$ of the solid. Intuitively, at length scales larger than elastocapillary length $l_\gamma$, 3D elasticity stabilizes the surface. What happens at ever smaller scales? In passive elastocapillarity, stability is provided by positive surface tension. By contrast, within active elastocapillarity, the destabilizing contribution of negative surface tension $\gamma$ must be regularized by higher-gradient stresses. We focus here on surface effects, and introduce a second \emph{bendoelastic} length $l_\kappa \equiv (\kappa/\mu)^{1/3}$. This length scale can arise, for example, from a surface with bending modulus $\kappa$~\cite{box_cloaking_2020}, or from the length dependence of active stresses $\gamma_\mathrm{a}$. A distinct stabilization mechanism is bulk dispersion. In the Methods, we consider the effects of a shear modulus $\mu(q)=\mu_0+\mu_1 q^2$, dependent now on the planar wavenumber $q$, which instead leads to a stabilizing lengthscale $\sqrt{\mu_1/\mu_0}$. We contrast this effect with viscous dissipation, from which no such stabilization is possible.
\begin{figure*}[t!]
\centering
\includegraphics[width=0.95\linewidth]{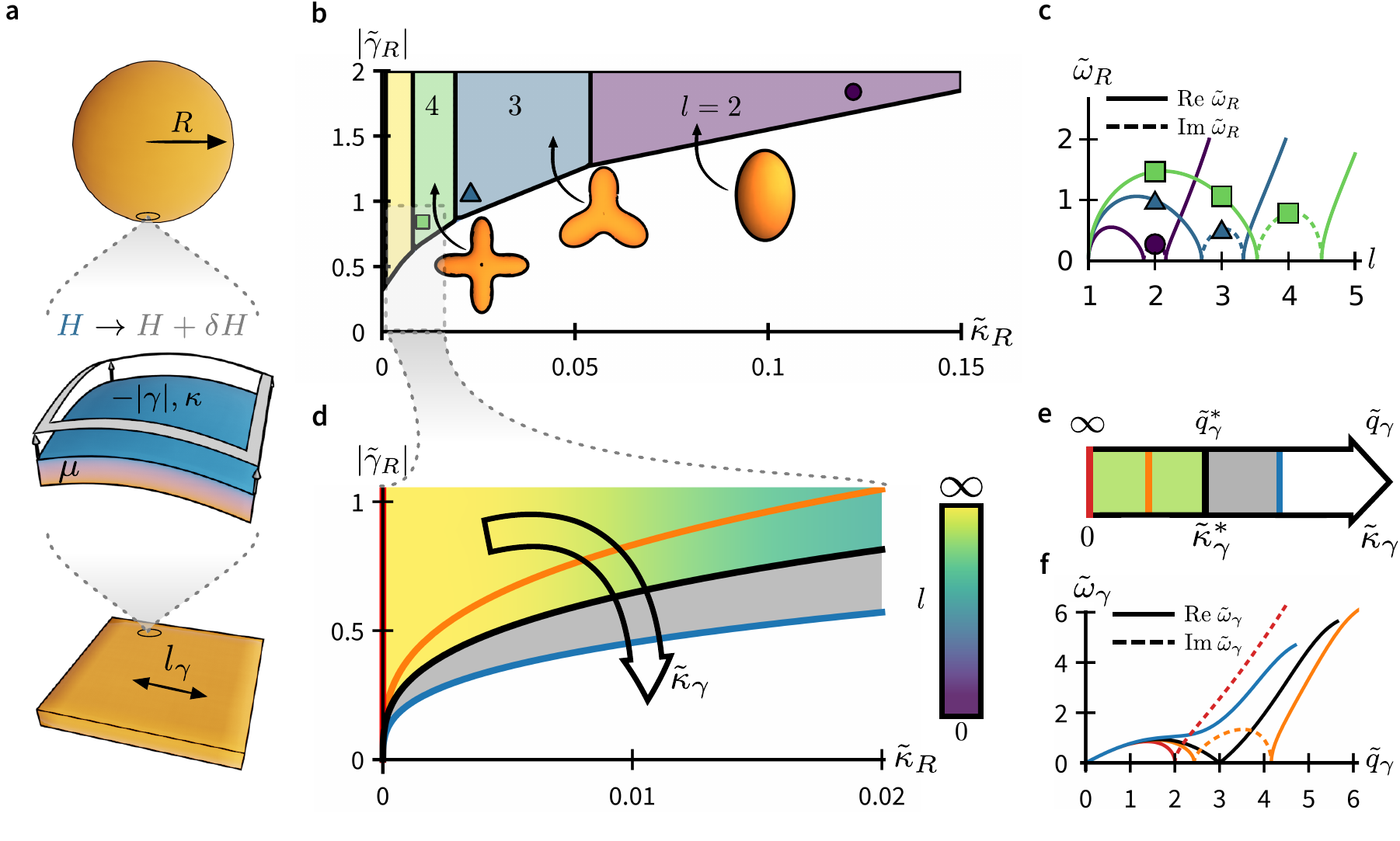}
    \caption{
      {\bf Shape instability and wave propagation within active elastocapillarity}. {\bf a.}~Schematic of an elastic solid with shear modulus $\mu$ and surface bending rigidity $\kappa$ deformed by a negative surface tension $-|\gamma|$, which leads to an excess mean curvature $H \rightarrow H+\delta H$. The stress-matching condition, Eq.~\ref{eq:StressMatch}, results in either restoring forces or surface instability and shape change. {\bf b.}~Phase diagram for the stability of an active elastocapillary sphere of radius $R$, controlled by rescaled surface tension $|{\tilde\gamma_R}|$ [$ \equiv |\gamma|/(\mu R)$, corresponding to activity] and bending modulus $\tilde{\kappa}_R$ [$\equiv \kappa/(\mu R^3)$]. The border between the white and the colored regions corresponds to activity strength at which the sphere goes unstable. The colors indicate the azimuthal mode number $l$, which is selected by the competition between shear and bending moduli. Insets show unstable modes at $l = 2, 3,$ and $4$. {\bf c.}~Dispersion $\tilde{\omega}_R \equiv \omega R\sqrt{\rho/\mu}$, with $\rho$ bulk density, of spherical oscillations corresponding to points marked by square, triangle, and circle in {\bf b}. A single mode is driven unstable ($\textrm{Im}~\tilde{\omega}_R > 0$, dashed line) as the threshold $\tilde{\gamma}_R$ is crossed. {\bf d--e.}~The limit $R \rightarrow \infty$ (equivalently, a zoom into the limit $l\rightarrow \infty$ from {\bf b}) asymptotically describes a half-space, and the phase diagram collapses to become one-dimensional ({\bf e}). The instability is controlled by the bending modulus $\tilde{\kappa}_\gamma \equiv \kappa \mu^2/|\gamma|^3$, given by the ratio of stabilizing elasticity ($\mu$ and $\kappa$) to destabilizing activity ($\gamma$). Each value of $\tilde{\kappa}_\gamma$ in {\bf e} (colored bars) gives a curve $\tilde{\gamma}_R \sim \tilde{\kappa}_R^{1/3}$ in {\bf d}. At $\tilde{\kappa}^*_\gamma=1/27$ [equivalently, $|\gamma^*| = 3(\kappa \mu^2)^{1/3}$], the half-space destabilizes at rescaled wavenumber $\tilde{q}^*_\gamma \equiv q |\gamma|/\mu =3$ ({\bf e}, green region). Even below threshold active driving a vestige of surface activity can be measured via the negative group velocity of surface elastocapillary waves ({\bf e}, gray region), which occurs for $\tilde{\kappa}_\gamma \lesssim 1.5\tilde{\kappa}^*_\gamma$ and $\tilde{q}_\gamma \gtrsim 0.8 \tilde{q}^*_\gamma$. {\bf f.}~Surface wave dispersions  $\tilde{\omega}_\gamma \equiv \omega \sqrt{\rho \gamma^2/\mu^3} $ corresponding to colored lines in parts {\bf d--e}, showing first a region of negative group velocity ($d \tilde{\omega}_\gamma/ d \tilde{q}_\gamma < 0$), and then instability ($\textrm{Im}~\tilde{\omega}_\gamma > 0$), developing as $\tilde{\kappa}_\gamma$ decreases.} \label{fig:lineartheory}
\end{figure*}

An object's shape results from the competition between bulk elasticity and boundary conditions containing active surface stresses. 
We solve the equations of linear elastodynamics with a stress-matching condition at the surface~\cite{tamim_elastic_2019,onodera_surface-wave_1998,harden_hydrodynamic_1991},
\begin{equation}
  -\sigma_{nn} = -2(\gamma-\kappa \nabla^2 )\delta H,
  \label{eq:StressMatch}
\end{equation}
for slow variations in initial curvature $H$, where $\sigma_{nn}$ is the component of the 3D elastic stress tensor normal to the surface, $\nabla^2$ is the covariant (surface) Laplacian, and $\delta H$ is the variation of mean curvature, see Fig.~\ref{fig:lineartheory}{\bf a}. In the Methods, we discuss the derivation of Eq.~\ref{eq:StressMatch} and details of our solutions.
Significantly, our approach accounts for 3D-elastic coupling between active surface elements, describing geometries inaccessible via a phenomenological free energy restricted to two dimensions, c.f.~Refs.~\cite{shlomovitz_exciting_2008,turlier_unveiling_2019} and Methods.

For any shape of size $R$, the solutions that we find must be characterized by the lengthscale triplet $(l_\gamma,l_\kappa,R)$. We then define two independent dimensionless ratios as the surface tension and bending modulus rescaled by the size $R$: $\tilde{\gamma}_R \equiv \gamma/(\mu R) = \mathrm{sign}(\gamma) l_\gamma/R$ and $\tilde{\kappa}_R \equiv \kappa/(\mu R^3) = (l_\kappa/R)^3$. 
When size $R$ is too large to be relevant (such as for an infinite half-space), the solutions depend only on the quantity $\tilde{\kappa}_\gamma =\kappa \mu^2/|\gamma|^3=(l_\kappa/l_\gamma)^3 $. Here $\tilde{\kappa}_\gamma$ describes the ratio of stabilizing elasticity ($\mu$ and $\kappa$) to destabilizing activity $|\gamma|$. We conclude that under overall rescaling, the phenomenology remains unchanged and that continuum elastocapillarity remains valid across any scale, from nanoparticles to macroscopic metamaterials.

\subsection{Tuneable Instabilities and Shape Selection}
Spheres minimize area at fixed volume, and positive surface tension drives every initial shape towards that of a sphere. By contrast, negative surface tension drives transitions away from a sphere into a variety of shapes. Our exact results demonstrate how to select between these shapes using the elasticity of the underlying solid. The phase diagram in Fig.~\ref{fig:lineartheory}{\bf b} shows that for low active driving, $|\tilde{\gamma}_R|$ is small and spherical shapes are stable. As activity increases, spheres spontaneously destabilize. The threshold activity for instability, encoded in $|\tilde{\gamma}_R|$, and the angular wavenumber $l$ of the dominant unstable mode are both determined by the balance of surface and 3D moduli $\tilde{\kappa}_R$ ($ \sim \kappa/\mu$).  

When the bending modulus is small compared to 3D elasticity (small $\tilde{\kappa}_R$), we find a wrinkling instability at wavelength $\sim l_{\kappa}$~\cite{sato_matsuo_kinetics_1988,li_surface_2011, fogle_shape_2013,tallinen_mechanics_2015,budday_size_2015,chang_extreme_2018,kusters_actin_2019}, allowing for control over fine structure and surface texture. However, as 3D elasticity weakens (large $\tilde{\kappa}_R$), both the wavelength and penetration depth of these wrinkles become comparable to the system size. In this limit, the instability manifests as bulk shape change. At the largest $\tilde{\kappa}_R$, we find shapes with uniaxial anisotropy, which are inaccessible, for example, via instabilities at fixed surface area~\cite{fogle_shape_2013}. These unstable modes originate in the dispersion relation, shown in Fig.~\ref{fig:lineartheory}{\bf c}, in which only a single mode is selected by negative surface tension [with corresponding points marked by square, triangle, and circle in Fig.~\ref{fig:lineartheory}{\bf b}]. This instability gives a scale-free tool for designing global shape change in bulk 3D solids, distinct from the buckling of thin elastic shells~\cite{quilliet_anisotropic_2008}.

For a flat object, energy injection will soften surface waves and drive surface wrinkling instabilities, Fig.~\ref{fig:lineartheory}{\bf d--e}. The two-dimensional phase diagram of a sphere collapses to the one-dimensional Fig.~\ref{fig:lineartheory}{\bf e}, described by $\tilde{\kappa}_\gamma$ and the planar wavenumber $\tilde{q}_\gamma \equiv q l_\gamma$. In Fig.~\ref{fig:lineartheory}{\bf f}, we show the planar dispersion relation $\tilde{\omega}_\gamma(\tilde{q}_\gamma)$, in which the frequency ${\omega}$ is rescaled by both elastocapillary length $l_\gamma$ and transverse speed of sound $c_T \equiv \sqrt{\mu/\rho}$. For weak active driving (large $\tilde{\kappa}_\gamma$) the half-space is stable, and plane waves stiffen at high wavenumbers. As active driving increases ($\tilde{\kappa}_\gamma$ decreases), energy injection softens high wavenumbers, leading first to negative group velocity $d \tilde{\omega}_\gamma/ d \tilde{q}_\gamma < 0$, and then to full-blown surface instability. Intuitively, this behavior stems from an effective shift of the shear modulus by negative surface tension, $\mu \rightarrow \mu -|\gamma|q$ (see Methods). This rescaling causes $\mu$, restoring elastic forces, and phase velocities all to vanish on a scale set by $l_\gamma$. The instability thresholds for wavenumber $q^* = l^{-1}_\kappa $ and active driving $|\gamma^*| = 3 (\kappa \mu^2)^{1/3}$ can both be tuned using the surface modulus $\kappa$ and 3D shear modulus $\mu$. In other words, by selecting the material parameters of the passive solid, we can select the first mode that goes unstable once activity is turned on.

\begin{figure*}[t!]
\centering
\includegraphics[width=0.99\linewidth]{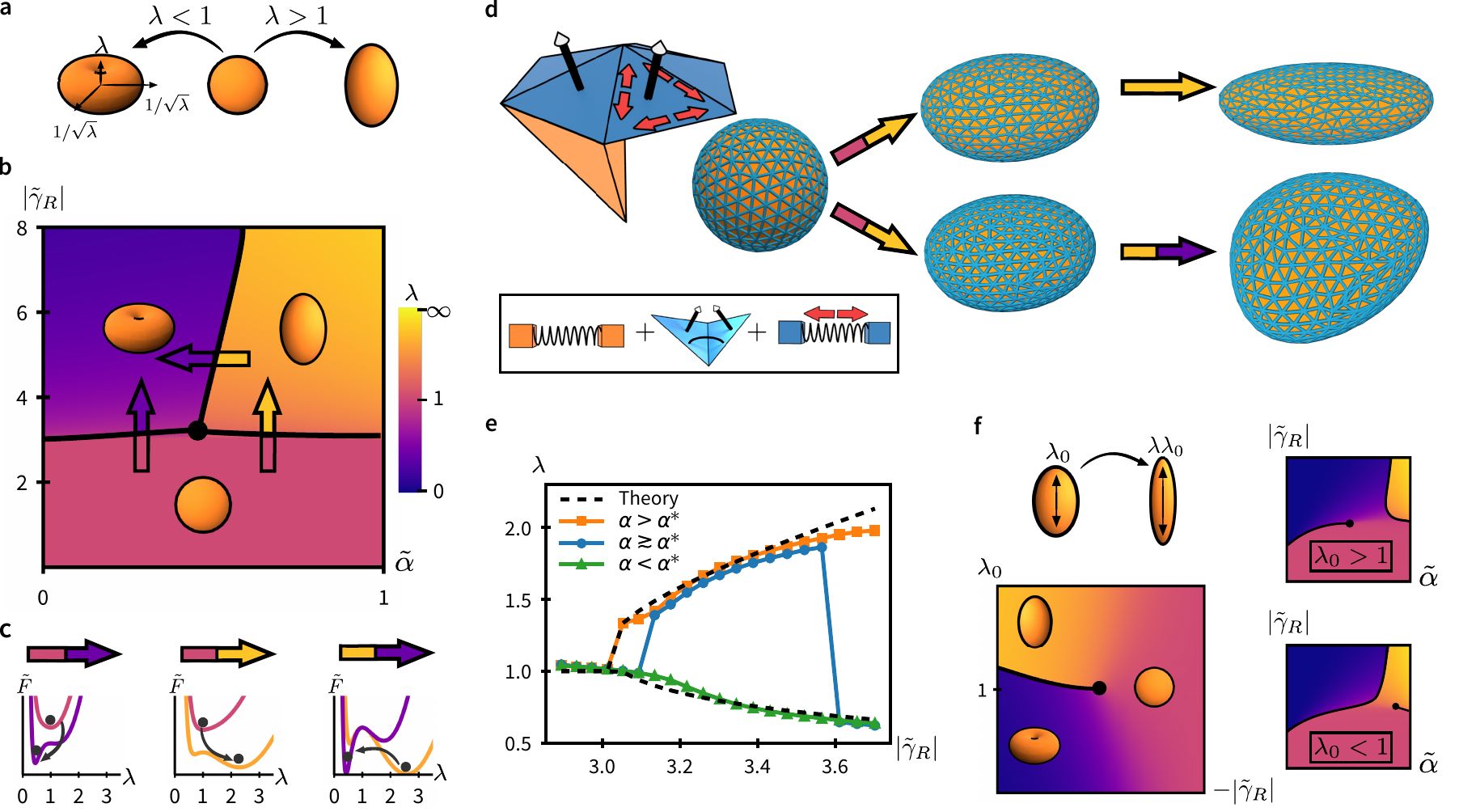}
    \caption{
      {\bf Nonlinear active elastocapillarity selects mode amplitude.} {\bf a.}~For large amplitudes, we describe a homogenous deformation by stretch factor $\lambda$, with $\lambda >1$ giving a prolate (worm-like) shape and $\lambda<1$, an oblate (pancake-like) one. {\bf b.}~Worm/pancake phase diagram. Beyond linear elasticity, sphere destabilization due to active driving $|\tilde{\gamma}_R|$ results in either a worm or a pancake, each stabilized by nonlinear terms. In the Mooney-Rivlin theory, the selected shape (mode amplitude) depends on material nonlinearity, parameterized by $\tilde{\alpha}$. In the neo-Hookean limit, $\tilde{\alpha}\rightarrow 0$, pancakes are favored. As $\tilde{\alpha}$ increases, material nonlinearity favors extension over compression, resulting in worms. These worms suffer a second `snap-through' transition into pancakes at higher active driving, even at fixed nonlinearity. These three discontinuous transition lines meet at a critical point. {\bf c.}~Landscapes for the effective energy describing discontinuous transitions along the arrows marked in {\bf b}. {\bf d--e.}~Simulations of a minimal ball-and-spring model demonstrate continuum theory predictions, see Supplementary Movie, and Methods for details. A spherical mesh of bulk nonlinear springs is coupled to surface springs exerting active stresses, with an energetic cost to bending deformations of surface plaquettes (inset). At a critical active driving the meshed solid destabilizes, with resulting shape tuneable via spring-level nonlinearity. For nonlinearity larger than the critical value ($\tilde{\alpha} > \tilde{\alpha}^*$, top), the mesh forms a worm-like structure, which elongates as active driving is further increased. For $\tilde{\alpha} \gtrsim \tilde{\alpha}^*$, the mesh destabilizes first to a worm, but then snaps through to a pancake upon increased active driving. {\bf e.}~Theory-simulation comparison for the dependence of stretch factor $\lambda$ on active driving $|\tilde{\gamma}_R|$.
      For numerical data, orange squares and blue triangles correspond to data shown in {\bf d}. Theoretical fit shows all of the (meta)-stable minima for the Mooney-Rivlin continuum theory Eq.~(\ref{eq:NonlinearFreeEnergy}) with parameters matched to the orange squares ($\tilde{\alpha} > \tilde{\alpha}^*$). {\bf f.}~If the initial shape is anisotropic, parameterized by initial stretch factor $\lambda_0$, the critical point from part {\bf b} splits. For both worm ($\lambda_0>1$, top right) and pancake ($\lambda_0<1$, bottom right), the initial anisotropy can either grow smoothly or snap through a transition line depending on path in parameter space. A cut in $|\tilde{\gamma}_R|$-$\lambda_0$ space through the critical point (bottom left) gives an Ising-like transition described by the Landau theory~(\ref{eq:LandauFreeEnergy}).}
    \label{fig:Nonlineartheory}
\end{figure*}

\subsection{Nonlinear Elasticity and Universality}
Linear analysis can select only mode number, not mode amplitude. We now explore the role of nonlinear bulk elasticity in selecting shape, focusing on the $l=2$ mode of uniaxial deformations. The amplitude of this uniaxial strain can be approximated by a homogenous deformation with principal stretch factor $\lambda$, see Fig.~\ref{fig:Nonlineartheory}{\bf a}. 

A general nonlinear elasticity introduces an infinite set of materials parameters, making inaccessible an exact solution like the one we obtained in the linear regime. In keeping with the minimal approach taken thus far, we first consider a simple model of nonlinear elasticity which accounts for the effects of material nonlinearity: the Mooney-Rivlin model, often used for rubbers and polymer gels such as those shown in Fig~\ref{fig:examples}{\bf b}--{\bf c}~\cite{treloar_physics}. We will return to the importance of this choice below. An effective energy $\tilde{F} \equiv F/\mu R^3$ for this far-from-equilibrium solid is given by 
\begin{equation}
  \tilde{F} = \tilde{F}_{\mathrm{NH}}+  \tilde{F}_{\mathrm{MR}}+\tilde{F}_{\mathrm{bend}}+ \tilde{\gamma}_R \tilde{A}.
  \label{eq:NonlinearFreeEnergy}
\end{equation}
For the active surface contribution $\tilde{\gamma}_R \tilde{A}$, we take the area $\tilde{A}$ of a uniaxial ellipsoid (Fig.~\ref{fig:Nonlineartheory}{\bf a}). This is balanced by a Helfrich bending energy $\tilde{F}_\mathrm{bend}$~\cite{zhong-can_bending_1989} and equilibrium bulk elasticity, composed of the neo-Hookean contribution $\tilde{F}_{\mathrm{NH}} \equiv (1-\tilde{\alpha})\left( 2 \lambda^{-1}+\lambda^2\right)/2$ and the Mooney-Rivlin term $\tilde{F}_{\mathrm{MR}} \equiv \tilde{\alpha}\left(\lambda^{-2}+2\lambda \right)/2$. We give expressions for the surface and bending energies in terms of $\lambda$ in the Methods. Crucially, this model includes a single parameter $\tilde{\alpha}$ to continuously tune material nonlinearity, which ranges between the neo-Hookean limit of only geometric nonlinearity, $\tilde{\alpha}=0$, and maximal nonlinearity, $\tilde{\alpha}=1$. 

Minimizing Eq.~\ref{eq:NonlinearFreeEnergy} yields the phase diagram shown in Fig.~\ref{fig:Nonlineartheory}{\bf b}. The discontinuous shape transitions indicated by solid lines correspond to bistable configurations in the effective energy $\tilde{F}(\lambda)$ (Fig.~\ref{fig:Nonlineartheory}{\bf c}). Increasing active driving destabilizes spheres, but now the elastic nonlinearity $\tilde{\alpha}$ controls whether the resulting shape is a prolate ellipsoid (a `worm') or an oblate one (a `pancake'). In the neo-Hookean limit $\tilde{\alpha}\rightarrow 0$, the preferred shape is a compressed pancake. Intuitively, this corresponds to maximizing surface area $\tilde{A}$ without any elastic effects (Fig.~\ref{fig:Nonlineartheory}{\bf c}, left panel). For sufficiently large $\tilde{\alpha}$, the preferred shape is instead an elongated worm (Fig.~\ref{fig:Nonlineartheory}{\bf c}, middle), reflecting the bias towards uniaxial elongation over compression encoded in the Mooney-Rivlin theory~\cite{treloar_physics}. These worms can undergo a second snap-through transition (Fig.~\ref{fig:Nonlineartheory}{\bf c}, right), morphing to pancakes as active driving $|\tilde{\gamma}_R|$ is further increased, or as the nonlinearity $\tilde{\alpha}$ is tuned.

In Figs.~\ref{fig:Nonlineartheory}{\bf d--e}, we realize these continuum predictions in a microscopic model of active elastocapillarity, as shown in the Supplementary Movie. We numerically simulate the deformation of a spherical mesh of bulk nonlinear springs, coupled to surface springs exerting active stresses, as shown in Fig.~\ref{fig:Nonlineartheory}{\bf d} (see Methods for details). At a critical active stress, the meshed solid destabilizes, exhibiting worms (Fig.~\ref{fig:Nonlineartheory}{\bf d}, top), pancakes, and snap-through transitions (Fig.~\ref{fig:Nonlineartheory}{\bf d}, bottom) depending on spring-level nonlinearity, in agreement with analytical predictions. In Fig.~\ref{fig:Nonlineartheory}{\bf e} we compare the continuum theory Eq.~(\ref{eq:NonlinearFreeEnergy}) with numerical data for the stretch factor $\lambda$. Significantly, the quantitative theory-simulation agreement near the transition points towards universality.

The three transition lines in Fig.~\ref{fig:Nonlineartheory}{\bf b} meet at a critical point. This critical point is a direct consequence of the symmetry of the initial shape. In contrast to a sphere, an initial uniaxial anisotropy $\lambda_0$ causes the critical point to split, Fig.~\ref{fig:Nonlineartheory}{\bf f}. For example, the phase diagram shown in Fig.~\ref{fig:Nonlineartheory}{\bf f}, top right, demonstrates how an initially elongated shape with $\lambda_0 >1$ can smoothly extend as active driving is increased. However, to compress into a pancake-like shape, the object must still cross a snap-through transition. The converse is true for initially compressed shapes (Fig.~\ref{fig:Nonlineartheory}{\bf f}, bottom right). In activity-anisotropy space, a cut through the critical point reveals an Ising-like transition (Fig.~\ref{fig:Nonlineartheory}{\bf f}, bottom left), with $-|\tilde{\gamma}_R|$ playing the role of temperature, and $\lambda_0$ an external field. This observation motivates a universal characterization of shape transitions near the critical point using Landau theory.

In the above discussion we introduced the Mooney-Rivlin model as a minimal choice. Other nonlinear elasticities will yield quantitatively different phase diagrams. However, near the critical point, the complete behavior of the active solid can be understood using symmetry-based arguments. For any initial shape, the Landau expansion guarantees that the effective free energy has the form
\begin{equation}
  \tilde{F}(\epsilon) = r \epsilon^2 - w \epsilon^3 + u\epsilon^4 - h \epsilon,
  \label{eq:LandauFreeEnergy}
\end{equation}
where the linearized strain $\epsilon$($ = \lambda -1$) plays the role of order parameter, the control parameter $r\sim \Delta \tilde{\gamma}_R$ probes the distance to linear instability, $w (\sim \Delta\tilde{\alpha}$ within the Mooney-Rivlin model) is the lowest-order nonlinear term, and $u > 0$ guarantees stability. The linear term $h\sim \lambda_0-1$ captures the effects of either shape anisotropy or external uniaxial stresses, and is absent for cubes, spheres, and other spherical tops (i.e., shapes with an isotropic moment-of-inertia tensor). As a result, a critical point is generically expected for these symmetric shapes, with three weakly discontinuous transitions emanating from it, as in Fig.~\ref{fig:Nonlineartheory}{\bf b}. Although Landau theory breaks down at higher strains, this critical point controls the entire phase-diagram shape.

We derive expressions for parameters $r$, $w$, $u$, and $h$ within the Mooney-Rivlin model in the Methods. However, we emphasize that the form of Eq.~\ref{eq:LandauFreeEnergy} is fully constrained by symmetry. As such, we expect the qualitative structure of our results to hold for a general nonlinear elasticity; the Landau theory presents a universal classification across all elastocapillary materials and shapes. The essential feature is simply that nonlinear effects appear in $w$, as is generically the case. In the Methods, we explore the phase diagram of the Gent model of rubber elasticity~\cite{gent_new_1996}, a singular example where nonlinearities only appear at quartic order.

\section{Discussion}
Active elastocapillarity couples field theories of different dimensionalities towards new materials design principles. Here we have adopted a continuum approach applicable across systems, using linear elasticity to describe shape selection and wave propagation analytically. We have found quantitative agreement between continuum nonlinear elasticity and particle-based numerics in predicting both the final shape of active elastocapillary solids and snap-through transitions between them. Landau theory explains this agreement in terms of universal behavior about a critical point, allowing a classification based solely on the initial symmetry of the solid.

We have focused on the minimal case of a passive elastic solid, coupled to active stresses in the form of an effectively negative surface tension $\Upsilon_{ij}=-|\gamma|\delta_{ij}$. The concept of active elastocapillarity may include a broader range of phenomena in which either bulk or surface stresses contain active components. Whilst our attention has been on synthetic realizations, some of these ideas may also be relevant across living systems where surface stresses affect shape. Examples of biological phenomena with surface growth and (2D or 3D) elasticity include cellular symmetry breaking~\cite{mietke_minimal_2019}, buckling of thin actin shells~\cite{kusters_actin_2019}, cellular layers~\cite{hannezo_theory_2014}, or tissues ~\cite{li_surface_2011,tallinen_mechanics_2015,budday_size_2015}, and modified wetting via differential growth~\cite{yousafzai_tissue_2020}. In contrast to the complexity of these systems, here we have highlighted how minimal ingredients such as isotropic surface stresses and bulk elastic nonlinearity can already be used to design surprising 3D metamaterial functionality.

The instabilities we have uncovered, from snap-through to smooth deformations, suggest active elastocapillarity as a portable mechanism to achieve complex reconfigurable shapes. At the macroscale, we envision the design of soft robotic arms composed of an elastic backbone covered in simple actuators (Fig.~\ref{fig:examples}{\bf c}). Scaling active elastocapillarity down to soft nanoparticles~(Fig.~\ref{fig:examples}{\bf b}), for which no reliable shape-control mechanism exists, may prove useful for applications ranging from drug delivery to self-assembly of photonic crystals.

\acknowledgements{\textbf{Funding}: J.B.~and A.S.~acknowledge the support of the Engineering and Physical Sciences Research Council (EPSRC) through New Investigator Award No.~EP/T000961/1. J.B.~and A.S.~acknowledge illuminating discussions throughout the virtual 2020 KITP program on ``Symmetry, Thermodynamics and Topology in Active Matter'', which was supported in part by the National Science Foundation under Grant No. NSF PHY-1748958.
This work was performed in part at the Aspen Center for Physics, which is supported by National Science Foundation grant PHY-1607611. The participation of A.S.~at the Aspen Center for Physics was supported by the Simons Foundation. T.R.W.~thanks the University of Birmingham for funding and support. \textbf{Author contributions}: J.B., T.R.W., and A.S. designed research; J.B. and A.S. developed the analytic theory; J.B. performed numerical simulations; J.B., T.R.W., and A.S. wrote the paper. \textbf{Competing interests}: The authors declare that they have no competing interests. \textbf{Data and materials availability}: All data needed to evaluate the conclusions in the paper are present in the paper and/or the Supplementary Materials. Code used to generate the simulation data reported in the manuscript is available at github.com/SouslovLab.}

\section{Methods}
\subsection{Feature sharpening}
\label{ssec:FeatureSharpening}
A positive surface tension minimises area, causing sharp edges and features to be rounded. Given a ridge~\cite{mora_softening_2015,jagota_surface-tension-induced_2012} or solid cone~\cite{ecole_dete_de_physique_theorique_les_houches_soft_2017} of material, positive surface tension blunts the tip to a smoothed cap. Here, we consider instead the effect of negative surface tension on a cone. We find that, by contrast, surface area maximisation sharpens the cone to a cusped structure. To show this, we first consider the elongation of each cylindrical slice of the cone, and then integrate along the conical axis to find the resulting shape.  

Refs.~\cite{mora_solid_2013, ecole_dete_de_physique_theorique_les_houches_soft_2017} consider the effect of surface tension $\gamma$ on a cylinder made of neo-Hookean material, of radius $\rho$ and shear modulus $\mu$. The total energy of the cylinder is  
\begin{equation}
E=E_{\mathrm{elastic}} + \gamma A,
\end{equation}
where $E_{\mathrm{elastic}} = 2\pi \mu \rho^2 \left[\left( \frac{1}{\lambda}-1\right)+\frac{1}{2}(\lambda^2 -1)\right]$ is the neoHookean elastic energy and $A=2\pi \rho \sqrt{\lambda}$ is the curved surface area of the cylinder (excluding the caps). Minimizing with respect to $\lambda$, we find that surface tension distorts the cylinder along its axis by a factor $\lambda$, where
\begin{equation}
    \lambda = \left[
    -\frac{\gamma}{4 \mu \rho} 
    + \sqrt{1 + \left(  \frac{\gamma}{4 \mu \rho}  \right)^2}
    \right]^\frac{2}{3}.
    \label{eq:CylinderStretch}
\end{equation}
This result applies regardless of the sign of $\gamma$. For a negative surface tension ($\gamma<0$), the stretch factor $\lambda$ is greater than one ($\lambda>1$), and the cylinder elongates. The volume is given by $\rho^2 \lambda$, and so for the incompressible neo-Hookean material there is a corresponding radial contraction to a deformed radius $R\sim\rho/\sqrt{\lambda}$. Taking the thin cylinder limit $|\gamma|/\mu \gg \rho$ of Eq.~\ref{eq:CylinderStretch}, for a negative surface tension we have a stretch factor $\lambda \sim (l_\gamma/\rho)^\frac{2}{3}$, where $l_\gamma = |\gamma|/\mu$ is the elastocapillary length.

Now, consider a three-dimensional cone of angle $\theta$ aligned along $z$ (Fig.~\ref{fig:examples}{\bf f}). Each cylindrical slice $z \rightarrow z +dz$ then experiences an elongation $\lambda(z) \sim (l_\gamma/\theta z)^\frac{2}{3}$. To find the total elongation, we integrate along the undeformed coordinate $z$ to obtain the deformed height $Z$:
\begin{equation}
    Z \sim  \int_0^z \lambda(z)dz \sim \left(\frac{l_\gamma}{\theta}\right)^{\frac{2}{3}} z^\frac{1}{3}.
    \label{eq:Zintermsofz}
\end{equation}
We can express this result in terms of the deformed radius $R$:
\begin{equation}
R(z)\sim \frac{\theta z}{\sqrt{\lambda(z)}} \sim l_\gamma \left(\frac{\theta z}{l_\gamma}\right)^\frac{4}{3}.
    \label{eq:rhointermsofz}
\end{equation}
Substituting Eq.~\ref{eq:rhointermsofz} into Eq.~\ref{eq:Zintermsofz} gives
\begin{equation}
    Z\sim \frac{l_\gamma}{\theta}\left(\frac{R}{l_\gamma}\right)^\frac{1}{4},
    \label{eq:cusp}
\end{equation}
where now $Z(R)$ is the height of the deformed cone as a function of its deformed radius. The exponent in Eq.~\ref{eq:cusp} is less than $1$, indicating a cusped structure, which becomes more pronounced with decreasing conical angle $\theta$.

\subsection{Modifications to active elastocapillarity: the effect of viscosity and bulk dispersion}
\label{ssec:Modifications}
\subsubsection{The effect of viscosity}
 In the main text, we have focused on inertial dynamics, but we may easily consider the effects of viscosity using a Kelvin-Voigt viscoelastic shear modulus,
\begin{equation}
    \mu(\omega)=\mu - i\nu\omega,
\end{equation}
in Eq.~\ref{eq:Dispersion}. For simplicity, we focus on the overdamped limit, neglecting inertia. In this limit, Eq.~\ref{eq:Dispersion} simplifies to
\begin{equation}
   i\nu \omega - \mu - \frac{\gamma}{2}q - \frac{\kappa}{2}q^3=0,
\end{equation}
or in dimensionless units
\begin{equation}
   i\tilde{\omega}_\nu - 1 + \frac{1}{2}\tilde{q}_\gamma - \frac{1}{2}\tilde{\kappa}_\gamma \tilde{q}_\gamma^3=0,
\end{equation}
where $\tilde{\omega}_\nu = \nu \omega/\mu$. Unlike some other examples of active solids \cite{scheibner_odd_2020}, here there is no phase lag between driving and response, so $\tilde{\omega}_\nu$ is always purely imaginary and we do not see overdamped waves. However, our tuneable shape instability remains, with the critical wavenumber and bending modulus as in Eq.~\ref{eq:qkcrit}. 

\subsubsection{Bulk dispersion} 
In the main text, we focused on a bending modulus as the regularizing mechanism for the $q\rightarrow \infty$ limit of active elastocapillarity. The lengthscale that $\kappa$ introduces, $l_\kappa$, comes from the surface physics. An alternate additional lengthscale, $l_\mu$, comes instead from stabilising higher-order gradients in the the bulk physics. In Fourier space, we consider a $q$-dependent 3D shear modulus
\begin{equation}
    \mu(q) = \mu_0 + \mu_1 q^2+...,
\end{equation}
where $l_\mu = \sqrt{\mu_1/\mu_0}$ and $\mu_1>0$. We can assess the effects of such higher-order terms using Eq.~\ref{eq:Dispersion}, by setting $\kappa=0$ and sending $\mu \rightarrow \mu+ \mu_1 q^2$. The result is a new dispersion,
\begin{equation}
    \rho \omega^2 - \frac{4\mu(q) q^2 \alpha_T}{q+\alpha_T}-\gamma q^3=0, 
    \label{eq:DispersionBulkDispersion}
\end{equation}
where now $\alpha_T= \sqrt{q^2 - \rho \omega^2/\mu(q)}$. In the absence of surface tension, Eq.~\ref{eq:DispersionBulkDispersion} admits Rayleigh waves $\omega \sim \sqrt{\mu(q)/\rho}q \sim \mu_1 q^2$) as $q\rightarrow\infty$. By contrast, pure capillary waves scale as $\omega \sim \gamma q^\frac{3}{2}$. A power count thus indicates that bulk dispersive effects also regularize the large $q$ limit. The asymptotic behaviour of Eq.~\ref{eq:DispersionBulkDispersion} as $q\rightarrow\infty$ is 
\begin{equation}
    \rho \omega^2  +|\gamma| q^3- \xi^2  \mu_1 q^4=0, 
      \label{eq:DispersionBulkDispersionAsymptotic}
\end{equation}
where $\xi= 0.955...$ is the ratio of Rayleigh to bulk wave velocity \cite{landau_theory_1986}. In dimensionless form, Eq.~\ref{eq:DispersionBulkDispersionAsymptotic} is
\begin{equation}
    \tilde{\omega}_\gamma^2  +\tilde{q}_\gamma^3- \xi^2 \tilde{\mu}_\gamma \tilde{q}_\gamma^4=0. 
\end{equation}
where $\tilde{\mu}_\gamma=(l_\mu/l_\gamma)^2$. Comparing Eq.~\ref{eq:DispersionBulkDispersionAsymptotic} to Eq.~\ref{eq:Dispersion}, we see the effects of bulk dispersion on high wavenumbers are qualitatively similar to a bending modulus, but the exact scalings differ. High wavenumbers are stabilised as $q^5$ with the bending modulus $\kappa$, but as $q^4$ with bulk dispersion $\mu_1$. Instead of $\tilde{\kappa}_\gamma = (l_\gamma/l_\gamma)^3$ controlling the phase planes Fig.~\ref{fig:lineartheory}{\bf d--e}, we have $\tilde{\mu}_\gamma=(l_\mu/l_\gamma)^2$. These dimensionless variables scale differently with their associated lengthscales $l_\kappa$ and $l_\mu$. One consequence of this difference would be a shifted scaling of the phase boundaries in Figs.~\ref{fig:lineartheory}{\bf d--e}. 

\subsection{Linear instability and shape change in an active elastocapillary droplet}
\label{ssec:Droplet}
In this section we derive the dispersion relation and linear instabilities of an active elastocapillary sphere. Recently, Ref.~\cite{tamim_elastic_2019} has given an analysis of the vibrations of the passive case, extending classical results for the purely elastic~\cite{eringen_elastodynamics} and capillary limits~\cite{landau_fluid_2013}. Here we instead consider the active case, in which the surface terms consist of both a negative surface tension $\gamma$ and bending modulus $\kappa$. The approach is to take a bulk ansatz satisfying the equations of linear elastodynamics~\cite{landau_theory_1986} and impose a matching boundary condition between bulk and surface stresses. This boundary condition leads to a solvability criterion, whose solution gives the dispersion and regimes of linear instability.

We first derive the boundary condition, balancing active stresses with restoring elasticity. Given a (two-dimensional) surface with surface tension $\gamma$ and bending rigidity $\kappa$, a variation of the Helfrich surface free energy
\begin{align}
    F = \int dA \left[
    2\kappa \left(H-c_0\right)^2+\gamma
    \right]
\end{align}
gives the shape equation for vesicles~\cite{zhong-can_bending_1989},
\begin{equation}
    P=  2 \kappa \left[\nabla^2 H + 2(H-c_0)(H^2-K+c_0 H) \right]- 2 \gamma H,
    \label{eq:ShapeEquation}
\end{equation}
where $P$ is (minus) the normal component of the bulk stress tensor, $H$ is the mean curvature of the solid's boundary, $K$ is the Gaussian curvature and $c_0$ allows for a preferred nonzero mean curvature. We now expand $H, K$ and $P$ to first order about a spherical shape, considering a normal perturbation $\psi \bf n$, where $\bf n$ is the outwards unit normal:
\begin{align}
  \label{eq:Perturbation1}
  \begin{split}
  H&=H_0 + \delta H + O(\psi^2) = -\frac{1}{R} + \delta H +O(\psi^2) ,  \\
  K&=K_0 + \delta K+ O(\psi^2)= \frac{1}{R^2} + \delta K + O(\psi^2),  \\
  P&=P_0 + \delta P+ O(\psi^2). 
  \end{split}
\end{align}
Here, $\delta H$, $\delta K$ are given by~\cite{capovilla_stresses_2002, capovilla_deformations_2003}
\begin{align}
  \label{eq:Perturbation2}
  \begin{split}
  \delta H = \frac{1}{R^2}\psi+\frac{1}{2}\nabla^2\psi, \\
  \delta K = -\frac{2}{R^3}\psi - \frac{1}{R}\nabla^2\psi.
  \end{split}
\end{align}
Substituting Eqs.~\ref{eq:Perturbation1} and \ref{eq:Perturbation2} into the shape equation Eq.~\ref{eq:ShapeEquation} yields the normal component of the stress matching condition
\begin{align}
    -\sigma_{nn} = -2\left[
    \left(\gamma + 2\kappa c_0\left(\frac{2}{R}+c_0\right) \right) 
    -\kappa \nabla^2
    \right] \delta H,
    \label{eq:SphericalStressMatch}
\end{align}
with the tangential component $\sigma_{\tau n} = 0$, where $\tau$ denotes the surface tangent. For an expansion in terms of spherical harmonics, we take $\psi =N Y_l^m = N P_l^m(\cos\theta) e^{i m \phi}$, with $P_l^m$ the associated Legendre polynomial, and $N$ a normalisation factor~\cite{jackson2007classical}. Then $\nabla^2 \psi = -l(l+1) \psi$ and Eq.~\ref{eq:SphericalStressMatch} simplifies to 
\begin{equation}
  -\sigma_{nn} = 
  \left[\gamma + 2\kappa c_0\left(\frac{2}{R}+c_0\right) + \frac{\kappa}{R^2}l(l+1) \right]\frac{(2-l(l+1))}{R^2}\psi.
  \label{eq:SphericalGammaShift}
\end{equation}
From Eqs.~\ref{eq:SphericalStressMatch}, \ref{eq:SphericalGammaShift} we see the effect of the bending modulus and spontaneous curvature is to shift $\gamma$ as
\begin{align}
\gamma \rightarrow \gamma + 2\kappa c_0\left(\frac{2}{R}+c_0\right) + \frac{\kappa}{R^2}l(l+1). 
\label{eq:SphericalGammaShift2}
\end{align}
Two natural special cases of this result are $c_0 =0$ (no spontaneous curvature) and $c_0 = -1/R$ (spontaneous curvature matching the initial mean curvature $H_0$). In the main text, we focus on the case $c_0 = 0$, for which Eq.~\ref{eq:SphericalStressMatch} simplifies to Eq.~\ref{eq:StressMatch}. However, note that only the last term in Eq.~\ref{eq:SphericalGammaShift2} depends on $l$ and the effect of $c_0$ can be completely reabsorbed into the effective surface tension $\gamma$.

With the mapping of Eq.~\ref{eq:SphericalGammaShift2}, the problem reduces to that considered in Ref.~\cite{tamim_elastic_2019}. The bulk stress tensor $\sigma$ is given by the general solution to the linear elastodynamic equations in spherical coordinates~\cite{eringen_elastodynamics}. Substituting this solution into Eq.~\ref{eq:SphericalGammaShift} gives the solvability condition, which we invert numerically to obtain the dispersion. Below, we state this solvability condition for the limit of an incompressible material, in terms of dimensionless variables coming from the sphere radius $R$ and associated timescale $\tau_R = R/\sqrt{\mu/\rho}$:
\begin{align}
\begin{split}
l &= q R, \\
\tilde{\omega}_R &= \tau_R \omega,  \\
\tilde{\gamma}_R &= \frac{l_\gamma}{R}, \\
\tilde{\kappa}_R &= \frac{\kappa}{ \mu R^3} =  \left(\frac{l_\kappa}{R}\right)^3.
\end{split}
\end{align}
In terms of these dimensionless quantities, the dispersion is given by the solution of
\begin{widetext}
\begin{multline}
              \tilde{\omega}_R\left[2+\tilde{\omega}_R^2-l^3\left(\tilde{\gamma}_R+\tilde{\kappa}_R l(l+1) \right)+2l\left(1+\tilde{\gamma}_R+\tilde{\kappa}_R l(l+1)
        \right)-l^2\left(4+\tilde{\gamma}_R+\tilde{\kappa}_R l(l+1) \right)\right]j_{l}(\tilde{\omega}_R)\\ 
              -2\left[\tilde{\omega}_R^2 + l\left(2+\tilde{\gamma}_R+\tilde{\kappa}_R l(l+1) \right)\left(2-l-l^2\right)\right]j_{l+1}(\tilde{\omega}_R)=0,
     \label{eq:DispersionSphere}
\end{multline}
\end{widetext}
where $j_l$ is a spherical Bessel function of the $l$th kind. 

Figs.~\ref{fig:lineartheory}{\bf b--c} are found by solving Eq.~\ref{eq:DispersionSphere} numerically. Note that the dispersion Eq.~\ref{eq:DispersionSphere} has an infinite number of branches, corresponding to the roots of $j_l$ --- spheroidal modes $\omega_R(s,l)$ are indexed by a radial `quantum number' $s$ and polar wavenumber $l$, being degenerate with respect to the azimuthal wavenumber $m$. Only the $s=1$ branch couples to the instability described in the main text, and it is this branch that is shown in Figs.~\ref{fig:lineartheory}{\bf b--c}.

\subsection{Waves and instabilities in an active elastocapillary half space}
\label{ssec:HalfSpace}
In this section, we derive the spectrum of the linearized equations of motion for an active elastocapillary half-space. These results follow from the $l\rightarrow \infty$ limit of section~\ref{ssec:Droplet}, in particular Eq.~\ref{eq:DispersionSphere}. However, an independent derivation in the planar case has the virtue of being much simpler than the spherical case, and we shall extend it to study the effects of viscosity and bulk dispersion in section~\ref{ssec:Modifications}. Passive elastocapillary waves have been studied from the perspective of a viscous fluid~\cite{harden_hydrodynamic_1991} or an elastic solid. Here, we take the elastic solids perspective~\cite{onodera_surface-wave_1998}, and consider the active case, in which a negative surface tension $\gamma$ is regularized in the high wavenumber limit by a bending modulus $\kappa$. Our approach applies equally to two- or three-dimensional materials. 

Before giving a detailed argument, basic scaling considerations capture the main phenomenology. Consider a slab of material of undeformed surface area $A$ with a deformed surface height $h\mathrm(x)$. The energy stored in surface deformations is $E_s \sim \left(\frac{\gamma}{2} h'(x)^2 + \frac{\kappa}{2} h''(x)^2\right) A$, and the bulk energy $E_b \sim~ \mu h'(x)^2 Al$, where $l$ is the depth that surface deformations penetrate into the bulk. Assuming $l\sim 1/q$, the total energy per unit volume is then $E/(lA) \sim (\mu + \frac{\gamma}{2} q + \frac{\kappa}{2} q^4 )h^2$, and we see that $\gamma$ acts as a $q$ dependent shift to the shear modulus, $\mu(q) = \mu + \frac{\gamma}{2} q$. As $q$ increases, for $\gamma<0$, $\mu(q)$ softens, with higher wavenumbers feeling progressively weaker elastic restoring forces. At $q \sim 1/l_\gamma$, restoring elasticity vanishes entirely, with $\kappa$ regularizing high wavenumbers. We thus expect the threshold wavenumber for instability to scale as $1/l_\gamma$.

 We now give a detailed derivation of the spectrum. We consider a half space $z\leq0$. The equations of linear elastodynamics in the bulk material are~\cite{landau_theory_1986}
 \begin{equation}
 \label{eq:elasticity}
    \rho \ddot{u_i} = \partial_j \sigma_{ij},
\end{equation}
 where $\rho$ is the density and $u_i$ is the displacement. For an isotropic material, the stress tensor $\sigma_{ij} = B \delta_{ij} u_{kk} + 2\mu\left(u_{ij}-\frac{1}{d}u_{kk}\delta_{ij}\right)$, where $u_{ij}= \frac{1}{2}(\partial_i u_j + \partial_j u_i)$ is the linearized strain, and $d$ is the spatial dimension. Equation \ref{eq:elasticity} supports longitudinal and transverse waves, of wavevector $q$ and frequency $\omega$, propagating along $x$ and decaying as $z\rightarrow -\infty$ \cite{landau_theory_1986, onodera_surface-wave_1998}:
\begin{equation}
\begin{aligned}
    {\bf u}_L = (q {\bf e}_x - i{\alpha_L}{\bf e}_z) \exp[i(q x - \omega t)+ \alpha_L z], \\
    {\bf u}_T = (i{\alpha_T} {\bf e}_x + q {\bf e}_z) \exp[i(q x - \omega t )+ \alpha_T z].
\end{aligned}
\end{equation}
Here $\alpha_T = \sqrt{q^2 - \rho \omega^2/\mu}$ and $\alpha_L = \sqrt{q^2 - \rho \omega^2/M}$ are the inverse penetration depths along $z$, with the longitudinal modulus $M = B + 2\frac{\mu (d-1)}{d}$. A general displacement ${\bf u} = u_x {\bf e}_x + u_z{\bf e}_z $ is written as
\begin{multline}
    {\bf u} = \big[  A_L (q {\bf e}_x - i{\alpha_L}{\bf e}_z)\exp{\alpha_L z} +  \\
    A_T(i{\alpha_T} {\bf e}_x + q {\bf e}_z)\exp{\alpha_T z}  \big] \exp{i(q x - \omega t)}.
    \label{eq:BulkAnsatz} 
\end{multline}

We now take the bulk ansatz Eq.~\ref{eq:BulkAnsatz} and substitute it into a stress matching boundary condition at $z=0$. The boundary has surface tension $\gamma$, bending modulus $\kappa$, and effective free energy
\begin{align}
    F = \int dx \left[\frac{\gamma}{2} \left(\frac{dh}{dx}\right)^2 + \frac{\kappa}{2}\left(\frac{d^2h}{dx^2}\right)^2 \right],
\end{align}
where $h(x)$ is the height of the free surface above $z=0$. Matching the $z$ component of bulk displacement $u_z$ to height $h$ yields the stress matching condition 
\begin{align}
\begin{aligned}
   \sigma_{zz}|_{z=0} &= \gamma \frac{d^2 u_z}{dz^2} - \kappa \frac{d^4 u_z}{dz^4},\\
   \sigma_{xx}|_{z=0}&=0.
\end{aligned}
\label{eq:NoStressPlanar}
\end{align}
Substituting Eqs.~\ref{eq:BulkAnsatz} into \ref{eq:NoStressPlanar} gives
\begin{widetext}
\begin{equation}
    \begin{bmatrix}
    [-i \alpha_L^2 M + i q^2 (M - 2\mu)]-i \alpha_L (\gamma+\kappa q^2) q^2& 2\mu q \alpha_T + (\gamma+\kappa q^2) q^3 \\
    2 \alpha_L q& i(\alpha_T^2 + q^2), \\
    \end{bmatrix}
     \begin{bmatrix}
    A_L\\
    A_T
    \end{bmatrix}
    = 0.
    \label{eq:Compressible}
\end{equation}
\end{widetext}
At this point we take the incompressible limit: as $B, M \rightarrow \infty$, $\alpha_L \rightarrow q$, and Eq.~\ref{eq:Compressible} simplifies to 
\begin{widetext}
\begin{equation}
    \begin{bmatrix}
    [ -i(2\mu q^2+ (\gamma+\kappa q^2) q^3)& 2\mu q \alpha_T + (\gamma+\kappa q^2) q^3 \\
    2 q^2& i(\alpha_T^2 + q^2) \\
    \end{bmatrix}
     \begin{bmatrix}
    A_L\\
    A_T
    \end{bmatrix}
    = 0.
    \label{eq:Incompressible}
\end{equation}
\end{widetext}
The dispersion is obtained from the solvability condition that the determinant of Eq.~\ref{eq:Incompressible} must vanish,
\begin{equation}
    \rho \omega^2 - \frac{4\mu q^2 \alpha_T}{q+\alpha_T}-\gamma q^3-\kappa q^5 =0. 
    \label{eq:Dispersion}
\end{equation}
Equation~\ref{eq:Dispersion} reduces to the passive elastocapillary dispersion (in the incompressible limit) for $\gamma > 0$ and $\kappa \rightarrow 0$~\cite{onodera_surface-wave_1998}, and the standard Rayleigh wave dispersion for $ \kappa, \gamma \rightarrow 0$~\cite{landau_theory_1986}. Intrinsic length and time scales are given by
\begin{align}
\label{eq:RelativeLengthScales}
\begin{split}
l_\gamma &= \frac{|\gamma|}{\mu}\quad (\mathrm{elastocapillary\ length)}, \\
l_\kappa &= \left(\frac{\kappa}{\mu}\right)^\frac{1}{3}\quad (\mathrm{bendoelastic\ length)}, \\
   \tau_\gamma &= \frac{l_\gamma}{\sqrt{\mu/\rho}} = \sqrt{\frac{\rho \gamma^2}{\mu^3}}
    \quad (\mathrm{elastocapillary\ time)},
\end{split}
\end{align}
which we use to define nondimensionalized variables
\begin{align}
\label{eq:nondim1}
\begin{split}
\tilde{q}_\gamma &= l_\gamma q, \\
\tilde{\omega}_\gamma &= \tau_\gamma \omega, \\
\tilde{\alpha}_\gamma &= \sqrt{\tilde{q}_\gamma^2 - \tilde{\omega}_\gamma^2},\\
\tilde{\kappa}_\gamma &= \left(\frac{l_\kappa}{l_\gamma}\right)^3 = \frac{\kappa \mu ^2}{|\gamma|^3}.
\end{split}
\end{align}
In dimensionless form, Eq.~\ref{eq:Dispersion} is then
\begin{equation}
    \tilde{\omega}_\gamma^2 - \frac{4\tilde{q}_\gamma^2 \tilde{\alpha}_\gamma}{\tilde{q}_\gamma+\tilde{\alpha}_\gamma}-\mathrm{sgn}(\gamma) \tilde{q}_\gamma^3-\tilde{\kappa}_\gamma \tilde{q}_\gamma^5 =0. 
    \label{eq:DispersionDimensionless}
\end{equation}
Here, we focus on the case of negative surface tension, $\mathrm{sgn}(\gamma)=-1$. The phase diagrams shown in Figs.~\ref{fig:lineartheory}{\bf d--e}, and the dispersions shown in Fig.~\ref{fig:lineartheory}{\bf f}, are found by solving Eq.~\ref{eq:DispersionDimensionless} numerically. However, the threshold at which instability occurs can be found analytically: letting $\tilde{\omega}_\gamma=0$ in Eq.~\ref{eq:DispersionDimensionless} gives the condition
\begin{equation}
\tilde{q}_\gamma^2(-\kappa \tilde{q}_\gamma^3 +\tilde{q}_\gamma -2 )=0.
\label{eq:zeros}
\end{equation}
We require the cubic in Eq.~\ref{eq:zeros} to posses degenerate roots, as must be the case at instability. This gives values for the wavenumber $\tilde{q}^*_\gamma$ and dimensionless bending modulus $\tilde{\kappa}^*_\gamma$ at which instability sets in:
\begin{align}
\label{eq:qkcrit}
\begin{split}
\tilde{q}^*_\gamma &= 3, \\
\tilde{\kappa}^*_\gamma &= \left(\frac{1}{3}\right)^3.
\end{split}
\end{align}
We may compare the structure of the above derivation to the intuitive argument presented at the beginning of this section. To the extent that $\alpha_T\approx q$ (strictly true in the Rayleigh wave limit $q\rightarrow 0$), Eq.~\ref{eq:Incompressible} shows that we can indeed formally absorb the effects of surface tension into a $q$-dependent shear modulus, $\mu(q)= \mu + \frac{\gamma}{2}q$. Further, referring to Eq.~\ref{eq:qkcrit}, we find that the threshold wavenumber for instability occurs at $q\sim1/l_\gamma$, as expected. 

\subsection{Explicit bulk elasticity versus an external surface potential}
\label{ssec:ExplicitVsExternal}
In this section, we contrast two approaches to modelling coupling between a surface and an elastic bulk. The first approach, taken in this work, explicitly uses both bulk and surface free energy terms. The free energy of a nearly-planar surface is then
\begin{equation}
    \begin{split}
    F &= \int d^3x \left[B u_{ii}^2 + \mu \left(u_{ik} - \delta_{ik} u_{ll}\right)^2 \right]  \\
    &+\int d^2 x \left[\frac{\gamma}{2} \left(\nabla h\right)^2 + \frac{\kappa}{2}\left(\nabla^2h\right)^2 \right].
    \end{split}
    \label{eq:approach1}
\end{equation}
Equation~\ref{eq:approach1} is translationally invariant (invariant under $u_i \rightarrow u_i+a,\ h \rightarrow h+a$). One consequence of this invariance is that the spectrum of Eq.~\ref{eq:approach1} is gapless, tending to the Rayleigh wave dispersion $\omega \sim q$. The scalings of the relevant length scales with material parameters (see Eq.~ \ref{eq:RelativeLengthScales}) are $l_\gamma =|\gamma|/\mu$ and $l_\kappa = (\kappa/\mu)^\frac{1}{3}$.

 The second approach is to use a surface-only free energy, with a phenomenological correction to account for bulk coupling. Such an approach has been used to analyse how the cytoskeleton in red blood cells affects their membrane fluctuations~\cite{gov_cytoskeleton_2003}, and has been proposed to model cytoskeletal wave propagation~\cite{shlomovitz_exciting_2008}. In this approach, we add an external Hookean potential $\frac{V_0}{2} h^2$ to a surface-only free energy:
\begin{equation}
    \begin{split}
    F = \int d^2x \left[\frac{V_0}{2} h^2+\frac{\gamma}{2} \left(\nabla h\right)^2 + \frac{\kappa}{2}\left(\nabla^2h\right)^2 \right].
    \end{split}
     \label{eq:approach2}
\end{equation}
Equation~\ref{eq:approach2} models the bulk elasticity as a fixed bed of springs at $h=0$, attached to the surface but not coupled to one another. Crucially, unlike Eq.~\ref{eq:approach1}, Eq.~\ref{eq:approach2} is not translationally invariant. This difference leads to qualitatively different predictions. For example, Eq.~\ref{eq:approach2} predicts a gapped spectrum, with a zero-frequency gap $\sim\sqrt{V_0}$. Further, Eq. \ref{eq:approach2} gives distinct scalings of $l_\gamma, l_\kappa$ with material parameters: $l_\gamma =(|\gamma|/V_0)^\frac{1}{2}$ and $l_\kappa = (\kappa/V_0)^\frac{1}{4}$.

\subsection{Nonlinear theory}
\label{ssec:NonlinearTheory}
\subsubsection{Derivation of the Mooney-Rivlin energy}
In this section we detail the derivation of Eq.~\ref{eq:NonlinearFreeEnergy}, giving expressions for $\tilde{F}_\mathrm{NH}$, $\tilde{F}_\mathrm{MR}$, $\tilde{\gamma}_R\tilde{A}$ and $\tilde{F}_\mathrm{bend}$ for the case of a uniaxial ellipsoid. We emphasize at the outset that the details of the results depend on the ellipsoidal geometry we have chosen, but their structure does not. One may repeat these calculations for other starting geometries, for example a cube, and obtain similar results.

Elastic deformations are described by the deformation gradient tensor $\Lambda$~\cite{warner_liquid_2003}. For a three-dimensional material, $\Lambda \equiv \frac{\partial\mathbf{X}}{\partial\mathbf{x}}$ is a three-dimensional tensor which gives the local mapping of material points from the undeformed state $\mathbf{x}$ to the deformed state $\mathbf{X}$. In general, $\Lambda$ depends on $\mathbf{x}$, the position within the undeformed state. Here we assume a homogeneous deformation, for which $\Lambda$ is constant. 
Given $\Lambda$, the three lowest-order rotational invariants which can appear in the elastic free energy density $f_{\mathrm{elastic}}$ are $I_1 = \mathrm{Tr} C$, $I_2 = \frac{1}{2}(\mathrm{Tr} C)^2 - \mathrm{Tr}(C^T C)$, $I_3 = \mathrm{Det}C$, where $C = \Lambda^T\Lambda$ is the right Cauchy--Green deformation tensor~\cite{warner_liquid_2003}. $I_1$ is the neo-Hookean term and $I_3$ describes volumetric deformations, i.e., $I_3=1$ for incompressible solids as we consider here. $I_2$ is the Mooney-Rivlin term, often used to phenomenologically account for material nonlinearity in rubbers \cite{treloar_physics}. The elastic part of the free energy density can be written as
\begin{equation}
  f_\mathrm{elastic}= c_1 I_1 + c_2 I_2+\cdots
  \label{eq:elastic1}
 \end{equation}
 We consider a uniaxial deformation, $\Lambda =\mathrm{diag}(1/\sqrt{\lambda},1/\sqrt{\lambda}, \lambda)$, and let $\mu = \frac{1}{2}(c_1+c_2)$ ($\mu$ is indeed the linear elastic shear modulus \cite{warner_liquid_2003}), $\alpha = \frac{1}{2}(c_1-c_2)$. Equation.~\ref{eq:elastic1} is then 
 \begin{equation}
  f_\mathrm{elastic}= \frac{\mu-\alpha}{2}\left( \frac{2}{\lambda}+\lambda^2\right) 
  + \frac{\alpha}{2}\left(\frac{1}{\lambda^{2}}+2\lambda \right). 
 \end{equation}
 For a uniaxial ellipsoid of radii $(R/\sqrt{\lambda},R/\sqrt{\lambda},R\lambda)$ the total elastic free energy is
\begin{equation}
  F_\mathrm{elastic}= \frac{4 \pi }{3} \mu \left[\frac{1-\tilde{\alpha}}{2}\left( \frac{2}{\lambda}+\lambda^2\right)
  + \frac{\tilde{\alpha}}{2}\left(\frac{1}{\lambda^{2}}+2\lambda \right)\right]R^3, 
  \label{eq:Fel}
 \end{equation}
 where we identify the first term in Eq.~\ref{eq:Fel} as $F_\mathrm{NH}$, and the second as $F_\mathrm{MR}$.
 
 The surface energy is 
\begin{equation}
\gamma A_\mathrm{ellipsoid}= \frac{2\pi \gamma}{\lambda}\left[1+\frac{\lambda^\frac{3}{2}}{e(\lambda)}\arcsin e(\lambda)\right] R^2,
\label{eq:AEllipsoid}
 \end{equation}
 where $e(\lambda)=\sqrt{1-\lambda^{-3}}$ is the eccentricity. For the bending energy $F_\mathrm{bend}$ we use the Helfrich form discussed in section~\ref{ssec:Droplet}: 
\begin{align}
    F_\mathrm{bend} = 2\kappa \int dA 
    \left(H-c_0\right)^2.
    \label{eq:Helfrich}
\end{align}
We now evaluate Eq.~\ref{eq:Helfrich} for the case of a uniaxial ellipsoid. Finite $c_0$ does not qualitatively change the structure of our results, and we consider $c_0=0$ for simplicity. The area element is 
\begin{align}
dA=\frac{R^2}{\sqrt{2}} \left(\frac{1}{\lambda^2} + \lambda + \left(\frac{1}{\lambda^2}-\lambda\right)\cos 2v\right)^\frac{1}{2}\sin v \ \mathrm{d}u\mathrm{d}v,
\end{align}
where $u$, $v$ are the azimuthal and polar angles on the ellipsoid. The mean curvature $H$ is 
\begin{align}
H=\frac{3+\lambda^3-(\lambda^3-1)\cos 2v}{\sqrt{2} R \lambda \left( \frac{1}{\lambda^2}+\lambda +\left(\frac{1}{\lambda^2}-\lambda\right)\cos 2v\right)^\frac{3}{2}}.
\end{align}
Equation \ref{eq:Helfrich} then simplifies to 
\begin{align}
    F_\mathrm{bend} = 4\pi\kappa \int_0^\pi \mathrm{d}v  \frac{\left(3+\lambda^3-(\lambda^3-1)\cos 2v\right)^2 \sin v}{2\sqrt{2}\lambda^2\left( \frac{1}{\lambda^2}+\lambda +\left(\frac{1}{\lambda^2}-\lambda\right)\cos 2v\right)^\frac{5}{2}},
\end{align}
which may be evaluated exactly; the result is 
\begin{align}
    F_\mathrm{bend} = 
   \frac{2\pi\kappa }{3} \left(\frac{2}{\lambda ^3}+\frac{3 \lambda ^3 \tanh ^{-1}\left(\sqrt{1-\lambda ^3}\right)}{\sqrt{1-\lambda ^3}}+7\right).
   \label{eq:Fbend}
\end{align}
Combining Eqs.~\ref{eq:Fel}, \ref{eq:AEllipsoid} and \ref{eq:Fbend}, all divided by $\mu R^3$, gives the total free energy Eq.~\ref{eq:NonlinearFreeEnergy}.
\subsubsection{Landau theory coefficients for the Mooney-Rivlin model}
In the main text, we argued that the behaviour of an active elastocapillary sphere near the critical point $\lambda=1$ can be understood based only on symmetries, using the Landau expansion Eq.~\ref{eq:LandauFreeEnergy}. Here, we derive the coefficients $r$, $w$, $u$ of Eq.~\ref{eq:LandauFreeEnergy} for the case of the Mooney-Rivlin free energy Eq.~\ref{eq:NonlinearFreeEnergy}. Expanding Eqs.~\ref{eq:Fel}, \ref{eq:AEllipsoid} and \ref{eq:Fbend} in the strain $\epsilon$ ($=\lambda -1$) we obtain
\begin{align}
\begin{split}
\frac{F}{\mu R^3} = A_2 \epsilon^2 + A_3 \epsilon ^3 + A_4 \epsilon ^4 + ... , \\
A_2 =\frac{2}{5} \pi  (4 \tilde{\gamma}_R +24 \tilde{\kappa}_R +5), \\
A_3=-\frac{4 \pi}{105} (35 \tilde{\alpha} +52 \tilde{\gamma}_R +360 \tilde{\kappa}_R +35), \\
A_4=\frac{2}{105} \pi  (105 \tilde{\alpha} +110 \tilde{\gamma}_R +1056 \tilde{\kappa}_R +70).
\end{split}
\label{eq:EllipsoidTaylor}
\end{align}
Equation.~\ref{eq:EllipsoidTaylor} has a critical point at $\tilde{\gamma}_R^*$, $\tilde{\alpha}^*$, $\tilde{\kappa}_R^*$, where 
\begin{align}
\begin{split}
\tilde{\gamma}_R^* = -\frac{1}{4} (5+24 \tilde{\kappa}_R^*), \\
\tilde{\alpha}^* = \frac{6}{35} (5- 8 \tilde{\kappa}_R^*). 
\end{split}
\label{eq:Location}
\end{align}
Expanding as $\Delta \gamma =\tilde{\gamma}_R-\tilde{\gamma}_R^*$, $\Delta \alpha =\tilde{\alpha}-\tilde{\alpha}^*$, $\Delta \kappa =\tilde{\kappa}_R-\tilde{\kappa}_R^*$, we obtain the structure of the free energy about this critical point:
\begin{align}
\begin{split}
\frac{F}{\mu R^3} = r \epsilon^2 + w \epsilon ^3 + u \epsilon ^4 + ... , \\
r=\frac{8}{5} \pi  (\Delta \gamma +6 \Delta \kappa), \\
u=\frac{1}{105} \pi  (-140 \Delta \alpha -208 \Delta \gamma -1440 \Delta \kappa ),\\
w=\frac{1}{105} \pi (504 \tilde{\kappa}_R^*+45),
\end{split}
\label{eq:EllipsoidLandau}
\end{align}
where we omit terms like $\Delta \gamma \epsilon^4$. 

\begin{figure}
\centering
\includegraphics{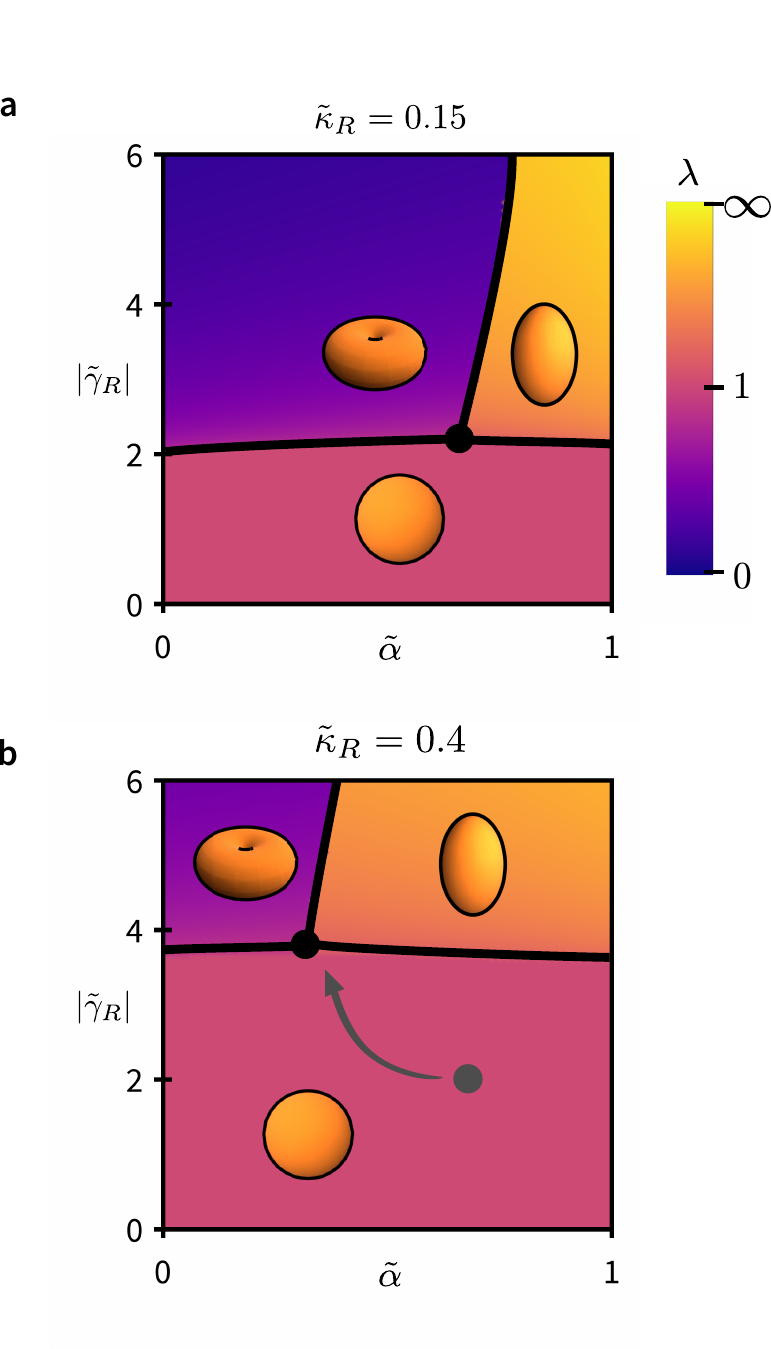}
    \caption{
      {\bf The structure of the worm/pancake phase diagram}. {\bf a.} Three weakly discontinuous transitions, described by the Landau theory Eq.~\ref{eq:EllipsoidLandau}, meet at a critical point given by Eq.~\ref{eq:Location}. {\bf b.} As the bending modulus $\tilde{\kappa}_R$ increases, the critical point is driven to larger active driving $|\tilde{\gamma}_R|$ and lower material nonlinearity $\tilde{\alpha}$, enlarging the worm-like region of parameter space.
    }
    \label{fig:VaryingKappa}
\end{figure}

In Fig.~\ref{fig:VaryingKappa} we show phase diagrams obtained from minimising the exact free energy Eq.~\ref{eq:NonlinearFreeEnergy}. We can interpret their structure in light of Eqs.~\ref{eq:Location}, \ref{eq:EllipsoidLandau}, with a focus on the novel interplay between bulk elasticity and surface effects. The Landau theory of Eq.~\ref{eq:EllipsoidLandau} corresponds to three weakly discontinuous transitions in the $\tilde{\alpha},|\tilde{\gamma}_R|$ plane, meeting at a critical point (Fig.~\ref{fig:VaryingKappa}{\bf a}). This structure is unchanged by varying the bending modulus $\tilde{\kappa}_R$. However, increasing the bending modulus drives the critical point to higher values of active driving $|\tilde{\gamma}_R|$ and lower material nonlinearity $\tilde{\alpha}$ (Fig~\ref{fig:VaryingKappa}{\bf b}), enlarging the region of phase space in which worms are favoured over pancakes. In this sense, both material nonlinearity $\tilde{\alpha}$ and bending rigidity $\tilde{\kappa}_R$ conspire to produce worm-like structures, as opposed to the more intuitively obvious pancake.

\subsection{Degenerate material nonlinearity: the Gent model}
In the main text, we explored the effects of material nonlinearity on the phase diagram of an active elastocapillary sphere, using the Mooney-Rivlin model as a minimal example. The Landau expansion for Mooney-Rivlin, Eq.~\ref{eq:EllipsoidLandau}, has the nonlinearity parameter $\tilde{\alpha}$ entering at cubic order in $\epsilon$, as is generically expected. The structure of the phase diagram Fig.~\ref{fig:Nonlineartheory}{\bf b} stems directly from this cubic term, and as such should be preserved across different choices of strain energy. Here, we investigate how this structure changes in the degenerate case in which nonlinearity only enters at quartic (or higher) order. A common nonlinear elastic model which exhibits this degeneracy is the Gent model~\cite{gent_new_1996}:
\begin{equation}
f_\mathrm{Gent}= -\frac{\mu}{2\beta}\log\left({1-\beta(I_1-3)}\right).
\label{eq:Gent}
\end{equation}
Gent elasticity is a correction to neo-Hookean behaviour at high extensions, used to model rubbers~\cite{gent_new_1996} and biological tissues~\cite{Bio_2017}. The material nonlinearity parameter $\beta$ models finite chain extensibility, with divergences in $f_\mathrm{Gent}$ occurring as $\lambda\rightarrow 0$ ($\lambda \sim \beta$) or $\lambda\rightarrow\infty$ ($\lambda\sim 1/\sqrt{\beta}$). Taking $\beta\rightarrow0$ gives neo-Hookean elasticity. 

Expanding Eq.~\ref{eq:Gent} as we did for the Mooney-Rivlin model Eq.~\ref{eq:EllipsoidTaylor} we obtain 
\begin{equation}
f_\mathrm{Gent}= \frac{3 \mu  \epsilon ^2}{2}
-\mu  \epsilon ^3
+\frac{1}{4} (9 \beta +4) \mu  \epsilon ^4 
+...,
\label{eq:GentExpansion}
\end{equation}
in which $\beta$ enters at quartic order. The result is that the Gent model behaves essentially as a neo-Hookean solid for all $\beta$.

\begin{figure}
\centering
\includegraphics{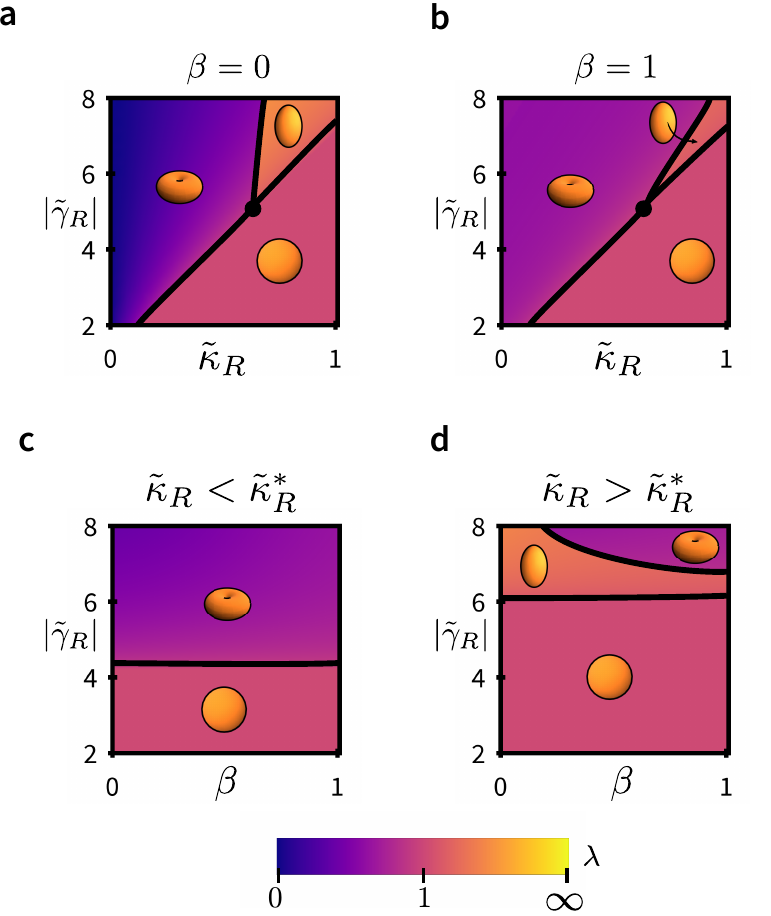}
    \caption{{\bf Phase diagram of the Gent model}. {\bf a--b.} Cuts in  $\tilde{\kappa}_R$--$|\tilde{\gamma}_R|$ space for the neo-Hookean limit $\beta=0$ ({\bf a}), and a generic nonzero $\beta$ ({\bf b}). The nonlinearity $\beta$ does not change the location of the critical point. {\bf c--d.} Cuts in $\beta$--$|\tilde{\gamma}_R|$ space above and below $\tilde{\kappa}^*_R$, showing a wormlike region opening up without a critical point.}
    \label{fig:PhasePlaneGent}
\end{figure}

Equation~\ref{eq:Location} gives the location of the critical point within the Mooney-Rivlin model. It describes a line in $\tilde{\alpha}$--$\tilde{\kappa}_R$--$|\tilde{\gamma}_R|$ space. This line intersects a generic coordinate plane to exhibit the critical point. In the Gent model, this line is parallel to the $\beta$ axis, and so a cut in material nonlinearity space will not exhibit a critical point. However, a generic cut, including the $\tilde{\kappa}_R$--$|\tilde{\gamma}_R|$ plane, will. In Fig.~\ref{fig:PhasePlaneGent}{\bf a--b}, we show the $\tilde{\kappa}_R$--$|\tilde{\gamma}_R|$ plane of the Gent phase diagram, for $\beta=0$ (the neo-Hookean limit) and a generic nonzero $\beta$. The phase plane contains a critical point, but upon varying $\beta$ its location does not change, as is generically expected. Rather, the wormlike region of the phase plane simply narrows. The location of the critical point can be found by setting $\tilde{\alpha}=0$ in Eq.~\ref{eq:Location}, giving $\tilde{\kappa}^*_R=5/8$. In Figs.~\ref{fig:PhasePlaneGent}{\bf c--d} we show the $\beta$--$|\tilde{\gamma}_R|$ plane below and above $\tilde{\kappa}^*_R$. The locus of the critical point runs parallel to these cuts, and we do not see a critical point in these diagrams. Instead, at low bending modulus, pancakes are favoured for all $\beta$. At high bending modulus, a wormlike region opens up, separated from pancakes by a curve running to $\beta\rightarrow\infty$. 

\subsection{Numerics} 
\label{ssec:Numerics}
In this section we describe the microscopic ball-spring model and numerical methods used to realise the predictions of the continuum theory shown in Fig.~\ref{fig:Nonlineartheory}.
\subsubsection{Microscopic Model}
\begin{figure}
\centering
\includegraphics{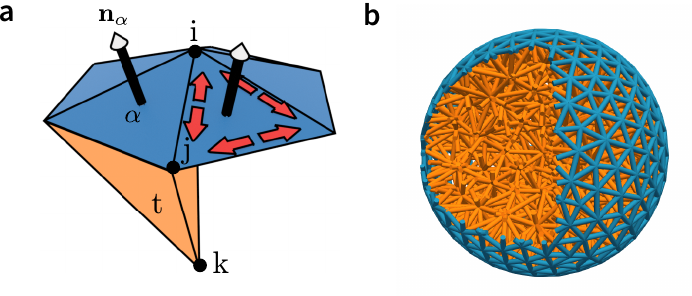}
    \caption{
      {\bf Microscopic ball-spring model}. {\bf a.} Schematic of the microscopic model at the surface of the meshed ball, showing vertices $i$, $j$, $k...$, edges $ij$, $jk...$, surface triangular plaquettes $\alpha$, and volume tetrahedra $t$. {\bf b.} Cut-through of the meshed ball used in simulation, showing pre-stressed surface springs (blue) and unstressed bulk springs (orange).
    }
    \label{fig:MicroscopicModel}
\end{figure}

We first construct a disordered tetrahedral meshing of the ball, as shown in Fig.~\ref{fig:MicroscopicModel}. We label the vertices $i$, edges $ij$, triangles $\alpha$ and tetrahedra $t$. A microscopic energy for deformations of this mesh is given by a spring energy $F_\mathrm{spring}$, a surface bending energy $F_\mathrm{bend}$, and an approximate volume constraint $F_\mathrm{vol}$:
\begin{equation}
    F=F_\mathrm{spring}+F_\mathrm{bend} + F_\mathrm{bulk}.
    \label{eq:MicroscopicEnergy}
\end{equation}
 For $F_\mathrm{spring}$, we place a spring along every edge $ij$ of the mesh. These springs have length $r_{ij}$, and rest length $r^0_{ij}$, from which we define the extension $\lambda_{ij} = r_{ij}/r^0_{ij}$. The spring energy is then
\begin{equation} 
F_\mathrm{spring}= \frac{k_\mathrm{m}}{2} \sum_{i>j} \left(1-{\alpha}_m\right)\left(2\lambda_{ij}^{-1} + \lambda_{ij}^2\right) 
+ {\alpha}_m\left(\lambda_{ij}^{-2} + 2\lambda_{ij}\right),
    \label{eq:MicroscopicEnergySpring}
\end{equation} 
i.e. each spring acts as an incompressible Mooney-Rivlin solid with microscopic neo-Hookean constant $k_m$ and material nonlinearity $\alpha_m$. To implement dilational surface stresses, we pre-stress the springs on the surface of the ball, initialising them at an extension $\lambda_{m}<1$. The bulk springs are initialised at their rest length, $\lambda=1$. We vary the macroscopic nonlinearity in material response $\tilde{\alpha}$ by varying $\alpha_m$ for the bulk springs, keeping the surface springs at $\alpha_m=0$. A bending energy $F_\mathrm{bend}$ is given by Ref.~\cite{boal_topology_1992}:
\begin{equation} 
F_\mathrm{bend} = \kappa_{m} \sum_{\alpha,\beta} (1- \mathbf{n}_\alpha\cdot\mathbf{n}_\beta),
\label{eq:DiscreteBending}
\end{equation} 
where the sum is over neighboring triangular plaquettes $\alpha, \beta$ on the surface of the ball, with $\bf{n}_\alpha$ the normal to plaquette $\alpha$. Finally, we approximately enforce incompressibility with an additional energetic penalty on volume changes of the tetrahedra $t$ of the mesh \cite{warner_liquid_2003}:
\begin{equation} 
F_\mathrm{vol} =B_{m} \sum_t (V_t-V^0_t)^2,
\label{eq:VolumeConstraint}
\end{equation} 
where $V_t$ is the current volume of a tetrahedron and $V^0_t$ is its rest volume.

The microscopic energy Eq.~\ref{eq:MicroscopicEnergy} contains $k_m$, $\alpha_m$ $\kappa_m$, $\lambda_m$ and $B_m$ as microscopic parameters. We now describe a mapping to the continuum shear modulus $\mu$, bulk modulus $B$, material nonlinearity $\tilde{\alpha}$, bending rigidity $\kappa$ and surface tension $\gamma$. Given a typical mesh lengthscale $a$, dimensional analysis gives
\begin{align}
\mu \sim \frac{k_\mathrm{m}}{a^3}, \\ 
\tilde{\alpha} \sim \alpha_m, \\
B \sim B_m a^3.
\label{eq:DimensionalAnalysis}
\end{align}
For an analytical estimate of the relation between $\kappa_m$ and $\kappa$, we may calibrate using the continuum limit of the discrete bending energy Eq.~\ref{eq:DiscreteBending} for a sphere, $4\pi \kappa_m/\sqrt{3}$~\cite{boal_topology_1992}. Comparing this to the continuum energy $8\pi \kappa$ gives the relation
\begin{align}
\kappa=\frac{1}{2\sqrt{3}}\kappa_m.
\label{eq:kappamtokappa}
\end{align}
For an estimate of the mapping from $\lambda_m$ to $|\gamma|$, one can show that the energy per unit area of a triangular spring mesh of side length $\lambda_m a$, composed of neo-Hookean springs, is given by
\begin{align}
|\gamma| = \frac{2\sqrt{3}}{a^2}k_m\left(1+ \frac{2}{\lambda_m^3}\right).
\label{eq:lambdamtogamma}
\end{align}

\subsubsection{Numerical Methods}
The data in Fig.~\ref{fig:Nonlineartheory} are generated by numerically minimizing Eq.~\ref{eq:MicroscopicEnergy} at fixed $k_m, \alpha_m, \kappa_m, B_m$, with $\lambda_m$ progressively decreasing from $\lambda_m=1$ (giving progressively stronger dilational surface stresses). The final state of each minimization is then used as an initialisation condition for the next. The minimizer used is the SciPy implementation of BFGS algorithm, with mesh vertex coordinates as input, and gradient norm stopping threshold of $10^{-2}$. For the data shown in Fig.~\ref{fig:Nonlineartheory}, a ball of radius $R=1$ (which defines the arbitrary spatial unit) is meshed with typical edge spacing $a=0.2$. Microscopic parameters $k_m =0.013$, $B_m = 50000$, $\kappa_m = 2.5$ are fixed for all runs. The three curves in Fig.~\ref{fig:Nonlineartheory}{\bf e} correspond to microscopic nonlinearities $\alpha_m= 0$ (green triangles), $0.3$ (blue circles), $0.4$ (orange squares). From the numerical data, we first map $\lambda_m$ to $\gamma$ using Eq.~\ref{eq:lambdamtogamma}. To obtain the values of $\lambda$ shown in Fig.~\ref{fig:Nonlineartheory}{\bf e}, an ellipsoid is then least-squares fit to the boundary vertices of the numerically relaxed mesh. The fit returns three ellipsoid axes, two of which are of similar magnitude ($\delta \lambda/\lambda < 0.1$), the third of which defines $\lambda$. 

Finally, we use the location of the critical point to fit the continuum theory Eq.~\ref{eq:NonlinearFreeEnergy} to this data, with $\tilde{\kappa}_R$, $\tilde{\alpha}$ as fitting parameters. The theoretical fit shown in Fig.~\ref{fig:Nonlineartheory} corresponds to $\tilde{\kappa}_R=0.3$, $\tilde{\alpha} = \tilde{\alpha}^*$ via Eq.~ \ref{eq:Location}. An independent assessment of $\tilde{\kappa}_R$ may be made using Eqs.~\ref{eq:kappamtokappa}--\ref{eq:DimensionalAnalysis}, from which we estimate $\kappa \approx 0.72$, $\mu \approx 5$ (independent measurement of $\mu$ for the meshed sphere used in simulation places $\mu \approx 4$). These estimates give $\tilde{\kappa}_R \approx 0.2$, consistent with the value we obtain from our fit.

\clearpage


\end{document}